\documentclass[aps,prd,eqsecnum,nofootinbib,showpacs]{revtex4-2}
\usepackage{amsmath,amssymb,bm,amsfonts}
\usepackage[dvips]{graphicx}
\usepackage{color}

\usepackage[utf8]{inputenc}
\usepackage{verbatim}

\numberwithin{equation}{section}

\usepackage{subfigure}
\graphicspath{{./fig_GL6/}}

\usepackage{cancel}

\newcommand{\be}{\begin{equation}}
\newcommand{\ee}{\end{equation}}
\newcommand{\bea}{\begin{eqnarray}}
\newcommand{\eea}{\end{eqnarray}}

\newcommand{\eq}[1]{Eq.~(\ref{eq:#1})}
\newcommand{\sect}[1]{Sec.~\ref{sec:#1}}
\newcommand{\appen}[1]{Appendix~\ref{sec:#1}}

\newcommand{\tabl}[1]{Table~\ref{table:#1}}

\newcommand{\del}{\partial}

\newcommand{\bra}{\langle}
\newcommand{\ket}{\rangle}

\newcommand{\eg}{{\it e.g.}}
\newcommand{\ie}{{\it i.e.}}

\newcommand{\bh}{black hole\ }

\newcommand{\SC}{superconductor}
\newcommand{\SCs}{superconductors}
\newcommand{\HSC}{holographic superconductor}
\newcommand{\HSCs}{holographic superconductors}

\bmdefine{\bmq}{{\bm{q}}}
\bmdefine{\bmk}{{\bm{k}}}
\bmdefine{\bmx}{{\bm{x}}}
\bmdefine{\bmy}{{\bm{y}}}
\bmdefine{\bmr}{{\bm{r}}}
\bmdefine{\bmnabla}{{\bm{\nabla}}}
\bmdefine{\bmA}{ \bm{A} }
\bmdefine{\bmD}{ \bm{D} }

\bmdefine{\bmPhi}{ \bm{\Phi} }
\bmdefine{\bmPsi}{ \bm{\Psi} }
\bmdefine{\bmcalO}{ \bm{\mathcal{O}} }
\bmdefine{\bmrho}{ \bm{\rho} }

\newcommand{\calR}{{\cal R}}

\newcommand{\vecx}{\vec{x}}

\newcommand{\epsmu}{\epsilon_{\mu}}

\newcommand{\nq}{\mathfrak{q}}

\newcommand{\nw}{\mathfrak{w}}

\newcommand{\calP}{{\cal P}}

\bmdefine{\bmg}{{\bm{g}}}
\bmdefine{\bmR}{{\bm{R}}}

\newcommand{\mfA}{\mathfrak{A}}

\newcommand{\half}{\frac{1}{2}}

\newcommand{\calA}{{\cal A}}

\newcommand{\calF}{{\cal F}}

\newcommand{\calJ}{{\cal J}}

\newcommand{\mass}{\text{M}}

\newcommand{\mfa}{\mathfrak{a}}

\newcommand{\eps}{\epsilon}

\newcommand{\SF}{superfluid}
\newcommand{\SFs}{superfluids}
\newcommand{\HSF}{holographic superfluid}
\newcommand{\HSFs}{holographic superfluids}

\newcommand{\dpsi}{\delta\psi}

\newcommand{\Jc}{J^*}

\newcommand{\calQ}{{\cal Q}}
\newcommand{\HSFHSC}{holographic superfluid/superconductor}
\newcommand{\HSFHSCs}{holographic superfluids/superconductors}

\newcommand{\itheta}{\theta_\eps}

\newcommand{\braket}[1]{\bra #1 \ket}
\newcommand{\dw}{\delta\omega}
\newcommand{\dq}{\delta q}
\newcommand{\dJ}{\delta J}
\newcommand{\dJt}{\braket{ \delta\calJ^t }}
\newcommand{\dJx}{\braket{ \delta\calJ^x }}
\newcommand{\Jte}{\calJ^t_\text{ext}}
\newcommand{\Jxe}{\calJ^x_\text{ext}}

\newcommand{\drho}{\braket{ \delta\rho }}

\newcommand{\vevpsi}{\braket{\dpsi}}
\newcommand{\vevpsic}{\braket{\dpsi^*}}

\newcommand{\dqs}{\delta(q^2)}

\newcommand{\sq}{\sqrt{2}}
\newcommand{\dP}{\delta P}
\newcommand{\dQ}{\delta Q}

\newcommand{\tilG}{\tilde{G}}

\newcommand{\Qt}{\calQ_{t\rho}}
\newcommand{\Qtp}{\calQ_{t\theta}}

\newcommand{\Qs}{\calQ_{\rho\rho}}
\newcommand{\Qsp}{\calQ_{\rho\theta}}
\newcommand{\Qi}{\calQ_{\theta\theta}}

\newcommand{\dmfA}{\delta\mfA}
\newcommand{\dcalA}{\delta\calA}

\newcommand{\dJr}{\delta J_+}
\newcommand{\dJi}{\delta J_-}

\newcommand{\Pole}{Pole }
\newcommand{\pole}{pole }

\begin{document}

%%%   Title Page   %%%

\title{Universal and nonuniversal pole-skipping points}
\author{Makoto Natsuume}
\email{makoto.natsuume@kek.jp}
\altaffiliation[Also at]{
Department of Particle and Nuclear Physics, 
SOKENDAI (The Graduate University for Advanced Studies), 1-1 Oho, 
Tsukuba, Ibaraki, 305-0801, Japan;
 Department of Mechanical Engineering, Mie University, 
Tsu, 514-8507, Japan.}
\affiliation{KEK Theory Center, Institute of Particle and Nuclear Studies, 
High Energy Accelerator Research Organization,
Tsukuba, Ibaraki, 305-0801, Japan}
%
%\arxivnumber{}
\preprint{KEK-TH-2883} 
\date{\today}
\begin{abstract}
We study the pole-skipping phenomenon for a \HSFHSC\ near the critical point using dual analytic Green's functions (arXiv:2605.20415 [hep-th]). (1) There is a hydrodynamic pole-skipping point that is common to all Green's functions: this point was obtained from the near-horizon analysis previously. (2) In addition, if one examines each Green's functions, the Green's functions have additional pole-skipping points. Unlike the conventional pole-skipping points, they do not satisfy the universality, \ie, they are not located at $\omega=-(2\pi T)ni$, where $n$ is an integer. (3) We also point out that a pole-skipping does not necessary implies an ambiguity of a Green's function. 
\end{abstract}
%
% KEK-TH-XXXX

%We study the pole-skipping phenomenon for a \HSFHSC\ near the critical point using an analytic solution we construct recently  (arXiv:2605.20415 [hep-th]). 
%In the pole-skipping phenomena, Green's functions are not uniquely determined at special points in the complex momentum space. We recently construct the analytic solution for a \HSFHSC\ near the critical point XX. 
%We study the pole-skipping phenomenon for a \HSFHSC\ near the critical point using analytic dual Green's functions
%solution we construct recently  
%(arXiv:2605.20415 [hep-th]). 
%The solution makes it possible to study the pole-skipping of the dual Green's functions explicitly.
% (1) There is a hydrodynamic pole-skipping point that are common to all Green's functions: this point was obtained from the near-horizon analysis previously (arXiv:2512.14883 [hep-th]). (2) In addition, if one examines each Green's functions, the Green's functions have additional pole-skipping points. Unlike the conventional pole-skipping points, they are not universal, \ie, they are not located at $\omega/(2\pi T)=-ni$, where $n$ is an integer. (3) We also point out that a pole-skipping does not necessary implies an ambiguity of a Green's function. 

%\tableofcontents

\maketitle

%%------------------
%\section{Time-dependence (gauge-invariant variables)}%\label{sec:}
%%------------------
%%%%%%%%%
\section{Introduction}%\label{sec:}
%%%%%%%%%

A pole-skipping point is a universal property of Green's functions \cite{Grozdanov:2017ajz,Blake:2018leo,Grozdanov:2019uhi,Blake:2019otz,Natsuume:2019xcy} found by the AdS/CFT duality or the holographic duality \cite{Maldacena:1997re,Witten:1998qj,Witten:1998zw,Gubser:1998bc}.%
\footnote{For a textbook, see, \eg, Refs.~\cite{CasalderreySolana:2011us,Natsuume:2014sfa,Ammon:2015wua,Zaanen:2015oix,Hartnoll:2016apf,Baggioli:2019rrs}} 
The location of pole-skipping points is universal in the complex momentum space. 
In this paper, we report the existence of pole-skipping points that do not satisfy the universality. To illustrate them, we use a \HSF/\HSC\ \cite{Gubser:2008px,Hartnoll:2008vx,Hartnoll:2008kx} as an example.

In AdS/CFT, one solves a bulk field equation and obtains  Green's functions in the dual theory [\eg, \eq{Green}].
Consider the field perturbation of the form $e^{-i\omega t+iqx}$, where $\omega$ is frequency and $q$ is wave number. Write a finite-temperature Green's function in the momentum space as 
%In the pole-skipping phenomena, finite-temperature Green's functions have the following property in the complex momentum space $(\omega,q)$: 
\begin{align}
G^R &= \frac{\calQ(\omega,q)}{\calP(\omega,q)}~.
%\label{eq:}
%
\end{align}
\begin{enumerate}
\item
At a pole-skipping point $(\omega_*,q_*)$,  $G^R=0/0$, namely the residue of a pole vanishes, or the pole is ``skipped."

\item
%Near a pole-skipping point, a Green's function $G^R$ typically takes the form
%\begin{align}
%
%G^R \propto \frac{ \delta\omega+\delta(q^2) }{ \delta\omega-\delta(q^2) }
%\label{eq:}
%
%\end{align}
Then, near a pole-skipping point, one can expand the Green's function as
\begin{align}
G^R &\sim \frac{ (\del_\omega \calQ)_* \dw + (\del_{q} \calQ)_* \dq +\cdots }{(\del_\omega \calP)_* \dw + (\del_{q} \calP)_* \dq +\cdots }
= 
 \frac{ (\del_\omega \calQ)_* + (\del_{q} \calQ)_* \frac{\dq}{\dw} +\cdots }{(\del_\omega \calP)_*  + (\del_{q} \calP)_* \frac{\dq}{\dw} +\cdots }
\label{eq:slope}
\end{align}
where ``$*$" denotes a value at the pole-skipping point $(\omega_*,q_*)$. 
Namely, the Green's function is not uniquely determined and depends on the slope $\dq/\dw$ how one approaches the pole-skipping point (``slope dependence"). %In the limit $\delta\omega,\delta(q^2)\to 0$, $G^R=0/0$, namely, the residue of a pole vanishes, or the pole is ``skipped."
\item
There is a universality for pole-skipping points. The pole-skipping point starts from
\begin{align}
\nw:=\frac{\omega}{2\pi T} =(s-1)i~, 
%\label{eq:}
%
\end{align}
and continues to $\nw=(s-1-n)i$ for a non-negative integer $n$ ($T$ is temperature and $s$ is the spin of the bulk field). 
For example, 
\begin{enumerate}
\item
For a scalar field, they are located at $\nw=-i, -2i,\cdots$. 
\item
For the Maxwell scalar mode $A_t,A_x$, they are located at  $\nw=0, -i, \cdots$. The $\nw=0$ point is a hydrodynamic pole typically. 
\item
For the gravitational sound mode such as $h_{tt}$, they are located at $\nw=+i, 0, -i \cdots$. It is argued that the $\nw = +i$ point is related to many-body quantum chaos \cite{Shenker:2013pqa,Roberts:2014isa,Roberts:2014ifa,Shenker:2014cwa,Maldacena:2015waa}. 
\end{enumerate}

\end{enumerate}
The pole-skipping phenomenon is analyzed from the near-horizon behavior of the bulk solution \cite{Grozdanov:2017ajz,Blake:2018leo,Grozdanov:2019uhi,Blake:2019otz,Natsuume:2019xcy}. This is a powerful method without constructing the full bulk solution. This also suggests that the universality of the pole-skipping phenomenon is related to the universal nature of the \bh horizon physics. 

However, this conventional formalism is applicable to a system with a single field. In general, one would like to study a system with multiple number of fields. In some cases, one can find a ``master variable," and the system reduces to a single field equation. Then, one can use the conventional formalism. However, it is in general very difficult to find a master variable. Its existence is not even guaranteed. 
Recently, we propose an alternative formalism that does not rely on the master variable \cite{Natsuume:2023lzy,Natsuume:2025nkx}. We then apply our formalism to \HSFs\ and find pole-skipping points for the system. 
%Thus, the pole-skipping phenomenon can now be studied for any system. 

On the other hand, the explicit dual Green's functions were rarely constructed except low-dimensional spacetime. We recently construct the analytic solution for a \HSFHSC\ %in the hydrodynamic limit 
near the critical point \cite{4th}, so it is possible to obtain the dual Green's functions (\sect{Green}) and study pole-skipping explicitly:
\begin{enumerate}
\item
There is a hydrodynamic pole-skipping point that are common to all Green's functions (\sect{universal}):
\begin{align}
(\omega_*, q^2_*)=\left(0,-\half\eps^2 \right)~,
%\label{eq:}
%
\end{align}
where $|\psi|=\eps$  is the spontaneous condensate .
This pole-skipping point has been obtained from the near-horizon analysis \cite{Natsuume:2025nkx}. We call such a point ``universal" pole-skipping point. 
\item
In addition, if one examines each Green's functions, each Green's functions have additional pole-skipping points (\sect{nonuniversal}). Unlike universal pole-skipping points, they are not located at $\omega/(2\pi T)=-ni$, where $n$ is an integer. We call such points ``nonuniversal" pole-skipping points. 
\item
Most nonuniversal pole-skipping points are located in the lower-half $\omega$-plane like the universal pole-skipping points for a Maxwell-scalar system. However, a nonuniversal pole-skipping point lies in the upper-half $\omega$-plane $\omega\propto i\eps^2$. The plane-wave ansatz $e^{-i\omega t}$ then seems to show an exponential growth.
\item
We also point out that a pole-skipping does not necessary implies an ambiguity of a Green's function (\sect{def}). 
\end{enumerate}
A pole-skipping does not necessary implies the ambiguity.  In \eq{slope}, we implicitly \textit{assume} that  $\del Q, \del P$ exist and they are different. This may not be true in general. For example, the pole and the residue may have the same form near a pole-skipping point and may become constant.

In \sect{further}, we examine some other examples to see the features of nonuniversal pole-skipping points:
\begin{enumerate}
\item
The ``minimal" \HSF\ with a background source for $\psi$.
\item
The ``nonminimal" \HSFs.
\item 
The generic model F universality class \cite{hohenberg_halperin}. Our system belongs to this universality class \cite{4th}. %part of which are our systems.
\end{enumerate}

The universal pole-skipping points can be obtained from the near-horizon analysis, but the analysis does not analyze the full solution. A ``nonuniversal" pole-skipping point arises only from the full solution. The origin of such a pole-skipping point is not entirely clear, but it is worthwhile to point out its existence. 

%%------------------
\section{The holographic superfluid/superconductor}\label{sec:preliminaries}
%%------------------

%%------------------
\subsection{Holographic \SF}%\label{sec:}
%%------------------

We consider the bulk 5-dimensional ``minimal" \HSFHSC:%
\footnote{We use upper-case Latin indices $M, N, \ldots$ for the 5-dimensional bulk spacetime coordinates and use Greek indices $\mu, \nu, \ldots$ for the 4-dimensional boundary coordinates. The boundary coordinates are written as  $x^\mu = (t, x^i) =(t, \vecx)=(t,x,y,z)$. $\omega$ and $q$ are collectively written as $k_\mu=(\omega,q,0,0)$.
}
\begin{subequations}
%\label{eq:}
\begin{align}
S_\text{bulk} &= \frac{1}{16\pi G_5}\int d^5x \sqrt{-g}(R-2\Lambda)+S_\text{m}~, \\
S_\text{m} &= -\frac{1}{g^2L} \int d^5x \sqrt{-g} \biggl\{ \frac{1}{4}F_{MN}^2 + L^2(|D_M\Psi|^2+m^2 |\Psi|^2) \biggr\}~, \\
F_{MN} &=\del_M A_N -\del_N A_M~, \quad
D_M =\nabla_M-iA_M~,\quad
\Lambda =-\frac{6}{L^2}~.
%\quad
%\Lambda =-\frac{6}{L^2}~.
%\label{eq:}
%
\end{align}
\end{subequations}
Here, we choose the mass dimensions as $[A_M]=\mass$, $[\Psi]=\mass^2$, and $[g]=\mass^0$. 
%We set $g=1$ below.
%The coupling $g$ has the dimension $[g]=\mass$, but we set $g=1$ below. 

We take the probe limit where the backreaction of the matter fields onto the geometry is ignored. %:
%\begin{align}
%
%\frac{1}{g^2N_c^2} \ll 1~,
%\label{eq:}
%
%\end{align}
%where $N_c$ is the number of ``colors" $N_c^2=8\pi^2L^3/(16\pi G_5)$, namely the probe limit corresponds to  the large-$N_c$ limit.
We set the bulk scalar charge $g=1$ below for simplicity.
Then, the background metric is given by the Schwarzschild-AdS$_5$ (SAdS$_5$) black hole:
\begin{subequations}
\label{eq:sads5}
\begin{align}
ds_5^2 &= \left( \frac{r}{L} \right)^2(-fdt^2+dx^2+dy^2+dz^2)+L^2\frac{dr^2}{r^2f} \\
&= \left( \frac{r_0}{L} \right)^2\frac{1}{u} (-fdt^2+dx^2+dy^2+dz^2)+L^2\frac{du^2}{4u^2f}~, \\
f &= 1-\left(\frac{r_0}{r}\right)^4 = 1-u^2~, 
%\label{eq:}
%
\end{align}
\end{subequations}
where $u:=r_0^2/r^2$.
The Hawking temperature is given by $\pi T =r_0/L^2$. For simplicity, we set the AdS radius $L=1$ and the horizon radius $r_0=1$, so we work in the unit $\pi T=1$. Then, $\nw:=\omega/(2\pi T)=\omega/2$.
The bulk matter equations are given by
\begin{subequations}
%\label{eq:}
\begin{align}
0 &= D^2\Psi-m^2\Psi~, \\
0 &= \nabla_NF^{MN} - J^M~,\\
J_M &= -i\{ \Psi^* D_M\Psi -\Psi(D_M\Psi)^*\} 
= 2\Im(\Psi^* D_M\Psi)~.
%\label{eq:}
%
\end{align}
\end{subequations}
In this paper, we consider the bulk scalar mass with $m^2=-4$. 
In the $A_u=0$ gauge, the $u\to 0$ asymptotic behaviors of matter fields are given by
%\margin{?}
\begin{subequations}
\label{eq:dictionary1}
\begin{align}
A_\mu &\sim \calA_\mu + A_\mu^{(+)} u~, \\
\Psi &\sim \frac{J}{2} u\ln u-\psi u~.
%\label{eq:dictionary1}
%
\end{align}
\end{subequations}
$\psi$ is the order parameter and $J$ is the source of the order parameter. 
Similarly, $\calA_t=\mu$ is the chemical potential, and and $\calA_i$ is the vector potential. 
%$\calJ^t$ is the charge density, and $\calJ^i$ is the current density.  
According to the standard AdS/CFT dictionary, the currents are given by%
\footnote{In the bulk 5-dimensions, one needs the counterterm action for the Maxwell field to cancel UV divergences. However, the scalar mode has no divergence at the leading order in our expansion, and the counterterm is not necessary.}
\begin{subequations}
%\label{eq:}
\begin{align}
\bra\calJ^t \ket &= -\sqrt{-g}F^{tu}|_{u=0}~,\\
\bra\calJ^x \ket &= -\sqrt{-g}F^{xu}|_{u=0}~.
%\label{eq:}
%
\end{align}
\end{subequations}
%Below we often omit brakets for simplicity.
%, but the counterterm makes no contribution at the leading order. 

%According to the standard AdS/CFT dictionary,
%\begin{subequations}
%\label{eq:dict}
%\begin{align}
%
%\Psi &\sim \frac{J}{2} u\ln u-\psi u~,\\
%A_t &\sim \calA_t - \frac{\calJ^t}{2} u~,\\
%A_i &\sim \calA_i + \frac{\calJ^i}{2} u~,
%\label{eq:}
%
%\end{align}
%\end{subequations}

We consider perturbations of the form $e^{-i\omega t+iqx}$. 
We consider perturbations around the background:
\begin{subequations}
%\label{eq:}
\begin{align}
\Psi &= \bmPsi+\delta\Psi~, \\
A_t &= \bmA_t+a_t~, \\
A_i &= 0 +a_i~,
%\label{eq:}
%
\end{align}
\end{subequations}
where boldface letters indicate the background solution.
In this paper, we consider the scalar mode that consists of $a_t,a_x,\delta\Psi,\delta\Psi^*$.

At high temperature, the background solution is given by
\begin{align}
\bmA_t = \mu(1-u)~, \quad
\bmA_i = 0~, \quad
\bmPsi =0~. 
\label{eq:sol_high}
\end{align}
Here, $\mu$ is the chemical potential. A \HSFHSC\ has 2 dimensionful control parameters $T$ and $\mu$, so the system is parameterized by a dimensionless parameter $\mu/T$. 
One can fix $T$ and vary $\mu$, or one can fix $\mu$ and vary $T$. We work in the unit $\pi T=1$, so we fix $T$.

The $\Psi=0$ solution becomes unstable at the critical point and is replaced by a $\Psi\neq0$ solution. When the background is the SAdS$_5$ black hole,  there exists a simple analytic solution at the critical point for $m^2=-4$ \cite{Herzog:2010vz}:
\begin{align}
\Psi \propto -\frac{u}{1+u}~,  \quad\text{at}\quad \left(\frac{\mu}{\pi T} \right)_c =2~.%\Delta=2~.
%\label{eq:herzog}
%
\end{align}
Below we utilize this solution to explore the system.

In the high-temperature phase, the bulk field $\delta\Psi$ decouples from the bulk Maxwell field $A_M$. In this case, the retarded Green's function associated with the order parameter $\psi$ is given by
\begin{align}
G^R_{\psi\psi} = -\frac{\del \psi}{\del J}~.
\label{eq:Green}
\end{align}
We obtain the bulk solutions in the low-temperature phase analytically and obtain Green's functions explicitly \cite{4th}. 

%\paragraph{Holographic \SC:}
%%------------------
\subsection{Holographic \SC}%\label{sec:}
%%------------------
So far, the boundary Maxwell field $\calA_\mu$ is added as an external source. Since the Maxwell field is not dynamical, one often calls this case the ``\HSF."  Traditionally, most literature call both \HSFs/\SCs\ as ``\HSCs" collectively, but we make the clear distinction between these two systems.
The distinction lies in the difference whether the boundary Maxwell field is dynamical or not.
%\margin{E OK}
\begin{enumerate}
\item
If the Maxwell field is dynamical, the system is a \HSC. 
\item
If the Maxwell field is added just as an external source, the system is  a \HSF. 
\end{enumerate}
Most literature actually study \HSFs.

To make the boundary Maxwell field dynamical, impose the ``holographic semiclassical equation"  as the boundary condition \cite{Natsuume:2022kic}:
\begin{align}
\del_\nu \calF^{\mu\nu} =e^2 \bra \calJ^\mu \ket~.
\label{eq:semiclassical}
\end{align}
Here, all quantities including the $U(1)$ coupling $e$ are the boundary ones.
%The vacuum magnetic permeability $\mu_0$ and the vacuum electric permittivity $\varepsilon_0$ are given by $\mu_0=1/\varepsilon_0=e^2$. 

Imposing the holographic semiclassical equation gives
\begin{subequations}
%\label{eq:}
\begin{align}
\del_i\calF^{ti} &= e^2 \bra\delta\calJ^t \ket +e^2\Jte~, \\
\to -\frac{1}{e^2}q^2 \dmfA_t &= \bra\delta\calJ^t \ket+\Jte~,
\label{eq:neumann}
\\
\dmfA_t &:= \dcalA_t+\frac{\omega}{q}\dcalA_x~.
%\to-\epse(k) q^2  \mfat &= \Jte~.
%\\
%\epse &:=\frac{1}{e^2} -\frac{K_L}{\omega^2}~.
%\label{eq:}
%
\end{align}
\end{subequations}
$\mfA_t$ is a gauge-invariant combination of the boundary Maxwell field. 

In this paper, we set $q\to lq, \omega\to l^2\omega$ and keep only the leading order terms in the $l$-expansion. In this expansion, the left-hand-side of \eq{neumann} is $O(l^2) \dmfA_t$ whereas the right-hand side is $O(l^0) \dmfA_t$.  Then, the left-hand side is negligible, and the problem reduces to the Neumann boundary condition problem at the leading order in our expansion:
\begin{align}
0= \bra\delta\calJ^t \ket+\Jte~, \quad
0= \bra\delta\calJ^x \ket+\Jxe~.
%\label{eq:}
%
\end{align}

%\paragraph{Counterterms:}
%In the bulk 5-dimensions, one needs the counterterm action for the Maxwell field to cancel UV divergences:
%\begin{align}
%
%S_\text{CT}= -\int d^4x\, 
%\frac{1}{4}
%\frac{1}{4g^2}
%\sqrt{-\gamma}\gamma^{\mu\nu}\gamma^{\rho\sigma}F_{\mu\rho}F_{\nu\sigma} 
%\times \ln\left(\frac{L}{r} \right)~,
%\label{eq:CT}
%
%\end{align}
%where $\gamma_{\mu\nu}$ is the 4-dimensional boundary metric (the 4-dimensional part of the bulk metric). 
%We use the coordinate $u$, and $(L/r)^2= (L/r_0)^2u$, so $\ln(L/r)=\ln(u^{1/2}L/r_0)$.
%Then,
%\begin{align}
%
%\bra \calJ^\mu \ket = \left. 2\del_u A_\mu
%- \del_\nu(\sqrt{-\gamma}F^{\mu\nu}) 
%\times \ln\left(\frac{L}{r_0}u^{1/2} \right) \right|_{u=0}~.
%\label{eq:}
%
%\end{align}
%At the leading order, the counterterm is necessary only for the Maxwell vector mode $\bra\calJ^y\ket$. 
%For $A_y\propto e^{-i\omega t+iqx}$,
%\begin{align}
%
%\bra \calJ^y \ket = \left. 
%2\del_u A_y
%+ 
%(\omega^2-q^2)\calA_y\left(\half\ln u-\ln \frac{r_0}{L} \right) \right|_{u=0}~.
%\label{eq:current_dict}
%
%\end{align}
%We assume $O(\omega)\sim O(q^2)$, so the $\omega^2$-term does not contribute at the leading order.

%%------------------
\subsection{Pole-skipping by the near-horizon analysis}%\label{sec:}
%%------------------
%\paragraph{Pole-skipping by the near-horizon analysis:}
%These nonuniversal pole-skipping points cannot be found from the standard near-horizon analysis. 
%Recall how usual pole-skipping points appear. 
The standard pole-skipping can be analyzed by the near-horizon analysis \cite{Grozdanov:2017ajz,Blake:2018leo,Grozdanov:2019uhi,Blake:2019otz,Natsuume:2019xcy}.
Let us consider a bulk field $\phi(u)$ and construct the solution starting from the horizon. 
The solution can be written as a power series in $(1-u)$:
\begin{subequations}
%\label{eq:}
\begin{align}
\phi(u) &=\sum_{n=0} \phi_n(1-u)^{n+\lambda}~,\\
\lambda_1 &= -i\omega/4~,
\lambda_2 = +i\omega/4~.
%\label{eq:}
%
\end{align}
\end{subequations}
The $\lambda_1 (\lambda_2)$ mode represents the incoming (outgoing) mode. We usually impose the incoming-wave boundary condition, so choose $\lambda_1$. The coefficient $\phi_n$ is obtained by a recursion relation. 

The situation changes %when $i\omega/2$ is an integer $n$. In this case, 
$\lambda_1$ and $\lambda_2$ differ by an integer. In such a case, the smaller root fails to produce the independent solution since the recursion relation breaks down at some $\phi_n$. Instead, the second solution would contain a $\ln(1-u)$ term and is not regular at $u=1$. 

However, this log term disappears for special values of $q$. Then, on has two regular solutions when $\lambda_2-\lambda_1=i\omega/2=n$ with appropriate $q_n$. This is the origin of a Green's function ambiguity at a pole-skipping point. 

This is a powerful method to find pole-skipping points because one needs to construct the near-horizon solution only. But clearly this method gives pole-skipping points only at $\omega=-2ni$. Our nonuniversal pole-skipping points do not arise in this way. The near-horizon solution is not the full solution, so nonuniversal pole-skipping points arise only from the full solution.

The hydrodynamic pole-skipping ($\omega=0$) and the chaotic pole-skipping ($\omega=+2i$) arise in a different way. In general, one has a set of the bulk field equations. In some cases, one can find a ``master variable." Then, the system reduces to a single field equation. These pole-skipping points arise because the master equation itself has the $0/0$ structure at pole-skipping points. 

However, it is unlikely that nonuniversal pole-skipping arise in this way because the master equation affects  all original field equations.

%%%%%%%%%
\section{Green's functions}\label{sec:Green}
% (using solution 2 v2)
%%%%%%%%%

In Ref.~\cite{4th}, we obtained the bulk solution near the critical point:
\begin{enumerate}
\item
We set 
$q\to lq, \omega \to l^2\omega, \eps \to l\eps$, 
and expands fields $a_t, a_x, \delta\Psi,\delta\Psi^*$
as a series in $l$. We also assign
\begin{align}
 a_t\sim O(l^2)~, a_x &\sim O(l)~,% \to \mfa_t\sim O(l^2)~, \\
%a_y &\sim O(l)~, a_u \sim O(l^2)~, \\
\delta\Psi \sim O(l)~. %\to \delta\rho \sim O(l)~, \delta\theta \sim O(l^0), \Theta\sim O(l)~.
%\label{eq:}
%
\end{align}
\item
For the boundary condition, we impose
\begin{enumerate}
\item
The incoming-wave boundary condition at the horizon $u=1$,
\item
$\mfa_t |_{u=0}=\dmfA_t$ where $\mfa_t:= a_t+\omega a_x/q$.
\end{enumerate}
\end{enumerate}
We obtain the bulk solutions at the leading order in the $l$-expansion and obtain the operator expectation values $\dJt,\dJx,\braket{\dpsi}, \braket{\dpsi^*}$ (\appen{original}). 

We also show that the result agrees with model F equations (\appen{F}). Namely, in the low-energy, in the long-wavelength limit, our \HSFHSC\ is described by model F equations.

%\newpage
%%------------------
\subsection{Holographic \SF}%\label{sec:}
%%------------------

%\paragraph{\HSF:}
We first rewrite our results in terms of gauge-invariant variables to avoid possible gauge ambiguities. We decompose $\psi$ as the amplitude and its phase:
\begin{subequations}
%\label{eq:}
\begin{align}
\psi &=\rho e^{i\theta} \to \delta\psi= \delta\rho+ i\eps \delta\theta~,\\
\delta \itheta &= \eps\left(\delta\theta-\frac{1}{iq}\delta\calA_x\right)~.
%\label{eq:}
%
\end{align}
\end{subequations}
$\delta\rho, \delta \itheta$ are gauge-invariant variables. Then,
\begin{subequations}
\label{eq:HSF_sol}
\begin{align}
\dJt &= 
2 q^2\frac{\calQ}{\calP} \dmfA_t 
+12 \eps \frac{\Qt}{\calP} \dJr
+ 4\eps \frac{\Qtp}{\calP} \dJi~, 
\quad O(l^2)
\\
\dJx &= 
\frac{\omega}{q} \left(2q^2\frac{\calQ}{\calP} \dmfA_t
+ 12\eps \frac{\Qt}{\calP}\dJr
+4\eps\frac{\Qtp}{\calP} \dJi \right) 
- \frac{\eps}{iq}\dJi~,
\quad O(l^3)
\\
\drho &= 
12\eps \frac{\Qt}{\calP} \dmfA_t 
+ 12 \frac{\Qs}{\calP}\dJr
+24\frac{\Qsp}{\calP}\dJi~, 
\quad O(l)
\\
\braket{\delta\itheta} &= 
-4\eps \frac{\Qtp}{\calP} \dmfA_t 
-24 \frac{\Qsp}{\calP}\dJr
+2\frac{\Qi}{\calP} \dJi~, 
\quad O(l)
%\label{eq:}
%
\end{align}
\end{subequations}
where $\dmfA_t = \dcalA_t+\omega \dcalA_x/q,  \dJ_+ =\dJ+\dJ^*,\dJi=(\dJ-\dJ^*)/i$.

The solution has a common denominator
\begin{align}
\calP &= 12q^6-48iq^4\omega-168q^2\omega^2+240i\omega^3+8q^4\epsilon^2-18iq^2\omega\epsilon^2-28\omega^2\epsilon^2+q^2\epsilon^4~,
%\label{eq:}
%
\end{align}
and there are 6 types of $O(l^4)$ numerators:
%\margin{TBC}
\begin{subequations}
%\label{eq:}
\begin{align}
\calQ &= 12q^4 - 24iq^2\omega -120\omega^2 + 20q^2\eps^2 - 74i\omega \eps^2 + 7\eps^4~,
\\
\Qt &= q^2(2q^2-8i\omega+\eps^2)~,
\\
\Qtp &= i\omega(24q^2-60i\omega+7\eps^2)~,
\\
%\calQ_x &=12q^4 -48i\omega q^2-72\omega^2+8q^2\eps^2-18i\omega\eps^2+\eps^4~,
%\\
\Qs &=2q^4-6iq^2\omega-4\omega^2+q^2\eps^2~,
\\
\Qsp &=i\omega(3q^2-6i\omega+\eps^2)~,
\\
\Qi &=12q^4-36iq^2\omega-24\omega^2+8q^2\eps^2-16i\omega\eps^2+\eps^4~.
%\label{eq:}
%
\end{align}
\end{subequations}
The current is conserved when there are no scalar source $\dJi$:
\begin{align}
\del_\mu\braket{\delta\calJ^\mu}= -\eps\dJi~,
\quad O(l^4)~.
%\label{eq:}
%
\end{align}

The Green's functions are given by
\begin{subequations}
%\label{eq:}
\begin{align}
-\dJt &= G^{tt} \dmfA_t + G^{t\rho}\dJr + G^{t\theta}\dJi~, \\
-\dJx &= G^{xt} \dmfA_t + G^{x\rho}\dJr + G^{x\theta}\dJi~,\\
-\drho &= G^{\rho t} \dmfA_t  + G^{\rho\rho}\dJr + G^{\rho\theta}\dJi~,\\
-\braket{\delta\itheta} &= G^{\theta t} \dmfA_t  + G^{\theta\rho}\dJr + G^{\theta\theta}\dJi~.
%\label{eq:}
%
\end{align}
\end{subequations}

%%------------------
\subsection{Holographic \SC}%\label{sec:}
%%------------------
%\paragraph{\HSC:}
For the \HSC, we impose the Neumann boundary condition  on the \SF\ solution \eqref{eq:HSF_sol}. 
In general, one can include the external sources $\Jte,\Jxe$.
So, we impose the Neumann boundary conditions $0=\dJt+\Jte,0=\dJx+\Jxe$. 
This gives $0 =  \eps\dJi +i\omega \Jte - iq \Jxe$, and 
%\margin{corrected v2}
%\margin{old convention}
\begin{subequations}
%\label{eq:}
\begin{align}
\dmfA_t &= -\frac{6\eps}{q^2}\frac{\Qt}{\calQ}\dJr 
- \frac{\calR_{tt}}{2\calQ} \Jte 
-\frac{2i}{q}\frac{\Qtp}{\calQ} \Jxe~, 
%+\frac{2\omega}{q}\frac{\calQ_t'}{\calQ} \Jxe~, 
\quad O(l^2)
% -\frac{\omega}{q}\dJr-\frac{i\eps}{q}\dJi + \Jxe~,
\\
\drho &= 12\frac{\calR_{\rho\rho}}{\calQ}\dJr 
+ \frac{\calR_{\rho t}}{\eps\calQ} \Jte  
-\frac{72\omega q}{\eps \calQ} \Jxe~, 
\quad O(l) 
\\
\braket{\delta\itheta} &=- \frac{72i\omega}{\calQ}\dJr
- \frac{8i\omega}{\eps}\frac{\calR_{\theta t}}{\calQ} \Jte 
+ \frac{2i}{q\eps}\frac{\calR_{\theta x}}{\calQ} \Jxe~, 
\quad O(l^3)
%\label{eq:}
%
\end{align}
\end{subequations}
Here,
%\margin{checked}
\begin{subequations}
%\label{eq:}
\begin{align}
\calR_{tt} &=12q^4 -48i\omega q^2-72\omega^2+8q^2\eps^2-18i\omega\eps^2+\eps^4~,\\
\calR_{\rho\rho} &= 2q^2-2i\omega+\eps^2~,\\
\calR_{\rho t} &= 72\omega^2-12q^2\eps^2+48i\omega\eps^2-6\eps^4~,\\
\calR_{\theta t} &= 3q^2-3i\omega-\eps^2~,\\
\calR_{\theta x} &=12q^4 - 12iq^2\omega +20q^2\eps^2 - 60i\omega \eps^2 + 7\eps^4~.
%\label{eq:}
%
\end{align}
\end{subequations}
The solution has a common denominator $\calQ$.
In the \SC\ case, the Maxwell field is dynamical, and the Goldstone boson ``disappears."  The Maxwell field becomes massive, and the longitudinal Maxwell mode has a hydrodynamic pole (\sect{pole_HSC}). One may write the results by $\dmfA_x:=-\delta\itheta$, but  we keep using the convention $\delta\itheta$. 

The Green's functions are given by
\begin{subequations}
%\label{eq:}
\begin{align}
-\dmfA_t &= A^{t\rho}\dJr 
+ A^{tt} \Jte 
+ A^{tx} \Jxe~, 
\\
-\drho &= \tilG^{\rho\rho}\dJr 
+ \tilG^{\rho t} \Jte 
+ \tilG^{\rho x} \Jxe~,
\\
-\braket{\delta\itheta} &=  \tilG^{\theta\rho}\dJr 
+ \tilG^{\theta t}\Jte
+ \tilG^{\theta x}\Jxe~.
%\label{eq:}
%
\end{align}
\end{subequations}
A Green's function is the ratio of the operator expectation value to a source. $A^{**}$ is not a Green's function, but it is a ``response" in the sense that the Maxwell field responds to a source, so we consider $A^{**}$  as well. For the \HSC, a Green's function is denoted as $\tilG^{**}$ to avoid the confusion from the $G^{**}$ for the \HSF.

%%------------------
\subsection{Pole structure}%\label{sec:}
%%------------------

The pole structure of Green's functions were discussed in Ref.~\cite{4th}, but we include it here for completeness.

%%------------------
\subsubsection{Holographic \SF}%\label{sec:}
%%------------------

Setting $\eps=0$ in \eq{bulk_Green} gives
\begin{subequations}
\label{eq:high}
\begin{align}
\dJt &= \frac{2q^2}{q^2-2i\omega} \dcalA_t~, \\
\vevpsi &= \frac{4}{q^2-(3+i)\omega} \dJ~, \\
\vevpsic &= \frac{4}{q^2-(3-i)\omega} \dJ^*~.
%\label{eq:}
%
\end{align}
\end{subequations}
Namely, in the high-temperature phase, there are 3 independent scalar modes $\dJt, \vevpsi,\vevpsic$, and each has a hydrodynamic pole. Also, $\dJt$ has a pole-skipping point $(\omega,q)=(0,0)$ whereas $\vevpsi,\vevpsic$ have no pole-skipping point. 
% [set $\eps=0$ in \eq{HSF_sol}].
%$\delta\calJ^t,\dpsi, \dpsi^*$. 

In the low-temperature phase, these modes are no longer independent, but one expects 3 poles that are related to 
$\dJt$, $\vevpsi$, $\vevpsic$ (\pole M,S,C, respectively). 
%$\dJt$ (\pole M), $\vevpsi$ (\pole S), and $\vevpsic$ (\pole C). 
In fact, the solution has common poles at
\begin{align}
\calP &= 12q^6-48iq^4\omega-168q^2\omega^2+240i\omega^3+8q^4\epsilon^2-18iq^2\omega\epsilon^2-28\omega^2\epsilon^2+q^2\epsilon^4 =0~.
%\omega= -\half i q^2-\frac{1}{4}i \epsilon^2~.
\label{eq:pole}
\end{align}
The pole $\calP$ is an $O(\omega^3)$ expression, so there are three  $\omega\neq0$ solutions in general. They correspond to \pole\ M, \pole S, and \pole C.
However, the exact expressions are rather involved. Schematically, we write
\begin{subequations}
\label{eq:pole_classify}
\begin{align}
\text{\Pole M: }  \omega&=\omega_M(q,\eps) \to  \dJt~. \\
\text{\Pole S: } \omega&=\omega_S(q,\eps)  \to \vevpsi~. \\
\text{\Pole C: } \omega&=\omega_C(q,\eps) \to \vevpsic~. 
%\label{eq:}
%
\end{align}
\end{subequations}
These poles are identified such that poles are smoothly connected between the high-temperature and the low-temperature phase at the critical point.
%\margin{E}
If one uses $w:=i\omega$, $\calP$ is a cubic equation in $w$ with real coefficients, so there is at least one real solution $w$. This corresponds to the \pole M. Namely, the pole of the Maxwell solution lies on the pure imaginary axis in the $\omega$-plane. 

One can obtain these poles approximately. This was first analyzed in Ref.~\cite{Herzog:2010vz}.
%\footnote{Note $\epsilon_\text{Herzog}^2 = 2\epsilon^2$. }
When $\eps \ll q$, 
\begin{subequations}
\label{eq:eps_small}
\begin{align}
\omega &= -\frac{i}{2} q^2-\frac{i}{10} \epsilon^2 +\cdots~,\quad(\text{M})~. \\
\omega &= \frac{3-i}{10}q^2+\frac{9-i}{120}\epsilon^2 +\cdots~, \quad(\text{S})~. \\
\omega &= \frac{-3-i}{10}q^2+\frac{-9-i}{120}\epsilon^2 +\cdots~, \quad(\text{C})~. 
%\label{eq:}
%
\end{align}
\end{subequations}
The first pole corresponds to the diffusive pole with an $\epsilon$-correction.
The last two poles correspond to the ones for order parameters.
For the order parameter poles, the coefficients of $q^2$ is complex, so the order parameters have damped oscillations. 

Similarly, when $\eps \gg q $, 
\begin{subequations}
\label{eq:q_small}
\begin{align}
\omega &= -\frac{7i}{60}\epsilon^2-\frac{89}{245}iq^2+\cdots~,
 \quad(\text{M})~.\\
 \omega &=+\frac{q\epsilon}{2\sqrt{7}}-\frac{33i}{196}q^2+\cdots~
 \quad(\text{S})~. \\
\omega &=-\frac{q\epsilon}{2\sqrt{7}}-\frac{33i}{196}q^2+\cdots~,
 \quad(\text{C})~.
%\label{eq:}
%
\end{align}
\end{subequations}
The last two poles have the linear dispersion and are proportional to $\epsilon$, so they are interpreted as the second sounds. The second terms of these poles represent the sound attenuation.

As one increases $\eps$, poles M,S, and C \eqref{eq:eps_small} become \eq{q_small}.

Note that these poles are approximate solutions in special limits, and they are not exact solutions of $\calP=0$. Thus, these dispersion relations are inappropriate to discuss pole-skipping points. 
The universal pole-skipping point belongs to \Pole M.

%It actually turns out that there is only 1 ``hydrodynamic" pole-skipping point (Pole M2) that are common to all Green's functions:
%\begin{align}
%
%(\omega,q^2)=\left(0,-\half \epsilon^2\right) \to \nq^2=-\frac{1}{8}\epsilon^2~.
%\label{eq:}
%
%\end{align}
%This agrees with our earlier result.
%Namely, some Green's functions diverge at the other poles.

%%------------------
\subsubsection{Holographic \SC}\label{sec:pole_HSC}
%%------------------

%The pole structure of Green's functions were discussed in Ref.~\cite{4th}, but we include it here for completeness.

The solution $\dmfA_t, \drho, \braket{\delta\itheta}$ has common poles at
 \begin{align}
\calQ &= 12q^4 - 24iq^2\omega -120\omega^2 + 20q^2\eps^2 - 74i\omega \eps^2 + 7\eps^4=0~.
%\label{eq:}
%
\end{align}
The pole $\calQ$ is an $O(\omega^2)$ expression, so there are two hydrodynamic poles instead of three for the \HSF. Again, write 
\begin{subequations}
%\label{eq:eps_small}
\begin{align}
\text{\Pole }A_x:   \omega&=\omega_A(q,\eps) \to  \braket{\delta\itheta}~. \\
\text{\Pole }\rho:  \omega&=\omega_\rho(q,\eps)  \to \drho~. 
%\label{eq:}
%
\end{align}
\end{subequations}

%They correspond to $\dmfA_x$ and $\dpsi_+$, and $\dmfA_t$ does not have a hydrodynamic pole.

One can obtain $\calQ=0$ poles approximately. 
When $\eps \ll q$, 
\begin{subequations}
%\label{eq:eps_small}
\begin{align}
\omega &= \frac{3-i}{10}q^2+\frac{21-37i}{120}\epsilon^2 +\cdots~,
 \quad(A_x) \\
\omega &= \frac{-3-i}{10}q^2+\frac{-21-37i}{120}\epsilon^2 +\cdots~,
 \quad(\rho) 
%\label{eq:}
%
\end{align}
\end{subequations}
Similarly, when $\eps \gg q $, 
\begin{subequations}
\label{eq:q_small_HSC}
\begin{align}
\omega &= -\frac{i}{2}\eps^2+\frac{4i}{23}q^2+\cdots~,
 \quad(A_x) 
\label{eq:HSC_scalar} \\
\omega &=-\frac{7i}{60}\eps^2-\frac{43i}{115}q^2+\cdots~,
 \quad(\rho) 
 %\label{eq:}
%
\end{align}
\end{subequations}
For the \HSF,  there are 2 damped oscillations. As the system is away from the critical point or $\eps$ becomes larger, they become second sound poles. For the \HSC, there are 2 damped oscillations again. However, as the system is away from the critical point, they become dissipative poles. 

The universal pole-skipping point belongs to \Pole $\rho$.

%%%%%%%%%
\section{Pole-skipping}%\label{sec:}
%%%%%%%%%

%%------------------
\subsection{Definition of pole-skipping}\label{sec:def}
%%------------------

A pole-skipping does not necessary implies the slope dependence of a Green's function. Write the Green's function as
\begin{align}
G^R &= \frac{\calQ(\omega,q)}{\calP(\omega,q)}~.
%\label{eq:}
%
\end{align}
Then, near a pole-skipping point,
\begin{align}
G^R &\sim \frac{ (\del_\omega \calQ)_* \dw + (\del_{q} \calQ)_* \dq +\cdots }{(\del_\omega \calP)_* \dw + (\del_{q} \calP)_* \dq +\cdots }
%\label{eq:}
%
\end{align}
where ``$*$" denotes a value at a pole-skipping point $(\omega_*,q_*)$. So, usually a pole-skipping implies the slope dependence $\dq/\dw$ of a Green's function. However,  this is the case when $\del Q, \del P$ exist and they are different. This may not be true, and there are actually 3 other cases:
%\begin{enumerate}[(a)]
\begin{itemize}
\item
Case 1:
The Green's function may be finite. This occurs when the pole and the residue have the same form near a pole-skipping point and cancel the slope dependence.
\item
Case 2:
The Green's function may vanish. This occurs when the order of a pole is smaller than the order of a zero, \eg, $\dw\dq/\dq\to 0$.
\item
Casse 3:
The Green's function still diverges but the order of a pole becomes smaller. This occurs when the order of a pole is larger than the order of a zero, \eg, $\dq/(\dw\dq) \to \infty$.

\end{itemize}
%\end{enumerate}
In any case, the order of a pole becomes smaller at a pole-skipping point. Thus, we define a pole-skipping point as follows:
\begin{center}
Definition of a pole-skipping point: \\
``The order of a pole becomes smaller at a pole-skipping point."
%``We define a pole-skipping\\ if the order of a pole becomes smaller at a pole-skipping point."
\end{center}

We classify pole-skipping points in 3 cases, but the classification may depend on how one approaches pole-skipping points. 
%\margin{E OK}
For example, the Maxwell diffusive Green's function typically takes the form 
\begin{align}
G^{tt} &=-\frac{q^2}{i\omega-Dq^2}~,
%\label{eq:}
%
\end{align}
with the diffusion constant $D$. If one expands $G^{tt}$ in terms of $\omega=0+\dw, q^2=0+\dqs$, one obtains the slope dependence:
\begin{align}
G^{tt} &\sim -\frac{\dqs}{i\dw-D\dqs}~.
%\label{eq:}
%
\end{align}
But if one expands $G^{tt}$ in terms of $\omega=0+\dw, q=0+\dq$,
\begin{align}
G^{tt} \sim- \frac{2q\delta q}{i\dw-Dq\dq}=0~.
%\label{eq:}
%
\end{align}
The Green's function now becomes Case 2.

%(nonuniversal, exceptional)

%%------------------
\subsection{Universal pole-skipping point}\label{sec:universal}
%%------------------
%$\mathfrak{G},\mathcal{G}$

Previously, we argued that our system has a hydrodynamic pole-skipping point at
\begin{align}
(\omega_*, q^2_*)=\left(0,-\half\eps^2 \right)
%\label{eq:}
%
\end{align}
from the near-horizon analysis \cite{Natsuume:2025nkx}. We first check that the dual Green's functions have this pole-skipping point explicitly. 

As a simple check of the universal pole-skipping point, the only solution of $\calP=q^2\calQ=\Qt=\Qtp=\Qs=\Qsp=\Qi=0$ is $(\omega,q^2)=(0,-\eps^2/2)$. 
%The solution also satisfies $2q^2-16i\omega+\eps^2=0$.

%%------------------
\subsubsection{Holographic \SF}%\label{sec:}
%%------------------
We expand Green's functions near the universal pole-skipping point.
%Let us look at the slope-dependence near the universal pole-skipping point. 
In the underlying bulk field equations, the variables $(\omega,q^2)$ naturally appear, so we expand Green's functions in terms of 
%$\omega=0+\delta\omega, q^2=-\eps^2/2+\dqs$:
\begin{align}
\omega=0+\delta\omega, q^2=-\half\eps^2+\dqs~.
%\label{eq:}
%
\end{align}
Then,
\begin{subequations}
%\label{eq:}
\begin{align}
\dJt &= 
 \frac{62i\dw-8\dqs}{ \dP} \eps^4\dmfA_t 
 + 12\frac{4i\dw-\dqs}{\dP} \eps^3\dJr 
 - \frac{20i\dw}{\dP} \eps^3\dJi~,
 \\
\dJx &= 
0\times \dmfA_t
+ 0\times\dJr
+\sq\dJi~,
\\
\drho &= 
 12\frac{4i\dw-\dqs}{\dP} \eps^3 \dmfA_t 
 + 12\frac{3i\dw-\dqs}{\dP} \eps^2\dJr 
 - \frac{12i\dw }{\dP} \eps^2\dJi~,
\\
\braket{\delta\itheta} &= \frac{20i\dw}{\dP} \eps^3 \dmfA_t 
  + \frac{12i\dw}{\dP} \eps^2\dJr 
  + \frac{4i\dw-8\dqs}{\dP} \eps^2\dJi~,
  \\
\dP &= \eps^4\{-3i\dw+2\dqs\}~.
%\label{eq:}
%
\end{align}
\end{subequations}
Then, all Green's functions have the universal pole-skipping point.
The slope-dependence does not appear for $\dJx$ 
because the corresponding Green's functions belong to Case 1 and 2.
%\margin{?}
%from the same reason as the pure Maxwell case as discussed earlier.

%%------------------
\subsubsection{Holographic \SC}%\label{sec:}
%%------------------
%To check the slope-dependence near the universal pole-skipping point, 
Similarly, expand Green's functions near the universal pole-skipping point:
%in terms of $\omega=0+\delta\omega, q^2=-\eps^2/2+\dqs$:
%\margin{checked}
\begin{subequations}
%\label{eq:}
\begin{align}
\dmfA_t &= 
 12\frac{4i\dw-\dqs}{\dQ} \eps\dJr 
 + \frac{-3i\dw+2\dqs}{\dQ} \eps^2 \Jte 
 + \frac{10\sq i \dw}{\dQ} \eps \Jxe~,
\\
\drho &= 
 24\frac{-i\dw+\dqs}{\dQ}\dJr
 +12 \frac{4i\dw-\dqs}{\dQ} \eps \Jte 
 - \frac{36\sq i \dw}{\dQ} \Jxe~,
\\
\braket{\delta\itheta} &= 
 \frac{-72i\dw}{\dQ}\dJr
 + \frac{20i\dw}{\dQ}\eps \Jte  
 -4\sq\frac{27i\dw-4\dqs}{\dQ} \Jxe~,
\\
\dQ &=2\eps^2\{-31i\dw+4\dqs\}~.
%\label{eq:}
%
\end{align}
\end{subequations}
 Again, all Green's functions have the universal pole-skipping point.
 
%%------------------
\subsection{Nonuniversal pole-skipping points}\label{sec:nonuniversal}
%%------------------

A nonuniversal pole-skipping point appears only from some Green's functions. Also, they do not appear at $\nw=-in$. We discuss such pole-skipping points in this section. 

Nonuniversal pole-skipping points are listed in \tabl{HSF} and \tabl{HSC}. (See \appen{slope}
% for the slope-dependence of 
for Green's functions near these pole-skipping points.) The \HSF\ has 3 sources, \ie, $\dmfA_t,\dJr,\dJi$. For example, consider the expectation value $\dJt$:
\begin{align}
-\dJt &= G^{tt} \dmfA_t + G^{t\rho}\dJr + G^{t\theta}\dJi~.
%\label{eq:}
%
\end{align}
\begin{enumerate}
\item
Some nonuniversal pole-skipping point appears for all Green's functions of the expectation value $\dJt$ (\eg, $G^{tt},G^{t\rho},G^{t\theta}$ for pole-skip~1).
\item
Some nonuniversal pole-skipping point appears only for specific Green's functions of the expectation value \eg, $G^{t\theta}$ for pole-skip~4. Namely,  they appear only when one turns off some sources ($\dmfA_t,\dJr$). Once one turns on these sources, the other Green's functions have poles there, not pole-skipping points, so these poles overwhelm the ambiguity of the nonuniversal pole-skipping point. 

\item
They all lie on the lower-half $\omega$-plane except ``pole-skip 5" for the \HSF. A pole-skipping point in the upper-half $\omega$-plane is unusual for a Maxwell-scalar system.
The gravitational sound mode and the Rarita-Schwinger field have universal pole-skipping points in the upper-half $\omega$-plane. Universal pole-skipping points are located at $\nw=(s-1)i$, so the sound mode pole-skipping point is $\nw=+i$. This shows an exponential growth and is related to many-body quantum chaos. The Rarita-Schwinger field has $s=3/2$, and the associated pole-skipping point is $\nw=+i/2$ \cite{Ceplak:2021efc}. The ``pole-skip 5" seems to show an exponential growth.

\item 
Most nonuniversal pole-skipping points $\omega_*$ are pure imaginary like universal pole-skipping points, but some nonuniversal pole-skipping points are complex (pole-skip 8 and 9 for the \HSC).
\item We use gauge-invariant variables and their sources for our analysis, but the choice of gauge-invariant variables is not unique. For example, one could use any linear combination of gauge-invariant sources $\dJr,\dJi$. This does not matter to the universal pole-skipping point because all Green's functions have the pole-skipping point. However, nonuniversal pole-skipping points appear in some Green's functions. Then, the change of gauge-invariant sources changes the location of nonuniversal pole-skipping points. However, nonuniversal pole-skipping points still exist.
% depends on how one chooses gauge-invariant variables.
\end{enumerate}

In AdS/CFT, one solves a bulk field equation, \eg, $\phi(u)$ and obtains the dual Green's function [\eg, \eq{Green}]. 
%The field equation \eqref{eq:scalar} has a regular singular point at the horizon $u=1$. 
In principle, one can construct the full solution by a power-series expansion:%
%\footnote{
%The power-series expansion is guaranteed to converge only inside a radius $\rho$ around $u=1$ in the complex $u$-plane, where $\rho$ is the distance to the nearest singular point. Because we construct the solution from $u=1$ to $u=0$, one needs $\rho \geq1$. The field equation has regular singular points at $u=0,\pm 1,\infty$, so the power-series expansion is guaranteed to converge at $u=0$.}
\begin{align}
\phi(u) = \sum_{n=0}\phi_n(u-1)^{n+\lambda}~.
%\label{eq:}
%
\end{align}
In a pole-skipping phenomenon, the bulk solution is not uniquely determined at pole-skipping points. As a result, the dual Green's function is not uniquely determined there.%
% v2.1
%\footnote{Because we solve a homogeneous linear differential equation, there is always an ambiguity of an overall factor of the bulk solution. But this is not relevant to Green's functions because one takes the ratio of the fast falloff to the slow falloff \eqref{eq:Green}.}

In general, it is difficult to obtain the closed form of the solution. But it was pointed out that the nonuniqueness comes from the near-horizon behavior of the solution (\sect{preliminaries}), so it is not necessary to obtain the full solution. It is enough to study the power-series solution up to $n$ to find the pole-skipping points at $\nw=-in$. 

In this sense, the conventional pole-skipping analysis does not see the full solution. So, each Green's function can have nonuniversal pole-skipping points other than universal pole-skipping points. On the other hand, this implies that nonuniversal pole-skipping points do not lack the universality because they are not related to the universal nature of \bh horizon physics.

%%============
\begin{table}[tb]
\begin{center}
\begin{tabular}{lccl}
	& $(\omega,q^2)$ &\Pole & Green's functions \\
	\hline
universal 	 & $\left(0,-\dfrac{1}{2}\eps^2 \right)$ & M & all \\
\hline
skip 0	&(0,0)& S,C & $\dJt, \dJx,\drho$ \\
skip 1	 & $\left(-\dfrac{7i}{60}\eps^2,0 \right)$ & M & $\dJt, \dJx$ \\%G^{t\rho}, G^{t\theta},G^{x\rho},G^{x\theta} $ \\
skip 2	 & $\left(-\dfrac{5i}{36}\eps^2,\dfrac{1}{18}\eps^2 \right) $ & M & $\dJt, \dJx$ \\%$ G^{t\rho}, G^{t\theta},G^{x\rho} $ \\
skip 3	 & $\left(-\dfrac{i}{12}\eps^2,-\dfrac{1}{6}\eps^2 \right) $ & M & $ G^{t\rho},G^{x\rho}, G^{\rho\rho}, G^{\rho\theta} $ \\
skip 4	 & $\left(0,-\dfrac{1}{6}\eps^2 \right)$ & M & $G^{t\theta},\dJx, G^{\rho\theta},G^{\theta\theta} $ \\
skip 5	 & $\left(+\dfrac{i}{12}\eps^2,-\dfrac{1}{2}\eps^2 \right) $ & C & $ G^{t\theta},G^{x\theta},G^{\rho\theta},G^{\theta\theta} $ 
\end{tabular}
\caption{Pole-skipping points for the \HSF.  
A pole-skipping point belongs to one of poles in \eq{pole_classify}. This is shown in the column ``Pole". 
%The column ``Pole" shows that a pole-skipping point belongs to a pole in the list \eqref{eq:pole_classify}. 
The last column shows the Green's functions in which the pole-skipping point appears. %\margin{E}
}
\label{table:HSF}
\end{center}
\end{table}
%%============
%%============
\begin{table}[tb]
\begin{center}
\begin{tabular}{lccl}
	& $(\omega,q^2)$ &\Pole & Green's functions \\
	\hline
universal 	 & $\left(0,-\dfrac{1}{2}\eps^2 \right)$ & $\rho$ & all \\
\hline
skip 1	 & $\left(-\dfrac{7i}{60}\eps^2,0 \right)$ & $\rho$ & 
$A^{tx}, \tilG^{\rho x}, \tilG^{\theta x}$ \\
skip 2	 & $\left(-\dfrac{5i}{36}\eps^2,\dfrac{1}{18}\eps^2 \right) $ & $\rho$ & $\dmfA_t$ \\
% $A^{t\rho},A^{tt}, A^{tx}$ \\
skip 6	 & $\left( 0,-\dfrac{7}{6}\eps^2 \right) $ & $\rho$ & 
$A^{tx}, \tilG^{\rho x}, \braket{\delta\itheta}$  \\
skip 7	 & $\left( -\dfrac{1}{2}\eps^2,0 \right) $ & $A_x$ &  $\drho$ \\
%$G^{\rho\rho},G^{\rho t}, G^{\rho x}$ \\
\hline
skip 8	 & $\left( \dfrac{-3i-\sqrt{11}}{12}\eps^2, \dfrac{7-\sqrt{11}i}{12}\eps^2 \right) $ & $A_x$ & 
$A^{tt}, \tilG^{\rho t}, \tilG^{\theta t}$ \\
skip 9	 & $\left( \dfrac{-3i+\sqrt{11}}{12}\eps^2, \dfrac{7+\sqrt{11}i}{12}\eps^2 \right) $ & $\rho $ & 
$A^{tt}, \tilG^{\rho t}, \tilG^{\theta t}$ 
\end{tabular}
\caption{Pole-skipping points for the \HSC. The last two pole-skipping points appear when $\Jte,\Jxe\neq0$.}
\label{table:HSC}
\end{center}
\end{table}
%%============

Some nonuniversal pole-skipping points appear from simple exact poles. 
For the \HSF, $\calP=0$ has simple solutions when  $\omega=0$ or $q=0$:
\begin{enumerate}

\item
When $\omega=0$, 
\begin{subequations}
%\label{eq:}
\begin{align}
\calP &= q^2(2q^2+\eps^2)(6q^2+\eps^2)~. \\
\to 
q^2 &=-\half \eps^2~,
 \quad(\text{universal skip})~. \\
 q^2 &=-\frac{1}{6} \eps^2~,
 \quad(\text{skip 4})~. \\
  q^2 &=0~,
   \quad(\text{skip 0})~.
%\label{eq:}
%
\end{align}
\end{subequations}

\item
When $q=0$,
%\margin{corrected 5/22}
\begin{subequations}
%\label{eq:}
\begin{align}
\calP &=4i\omega^2(60\omega+7i\eps^2)~. \\
\to
\omega &= 0~,\\
\omega &= -\frac{7i}{60}\eps^2~, 
 \quad(\text{skip 1})~. 
%\label{eq:}
%
\end{align}
\end{subequations}

\end{enumerate}

For the \HSC, $\calQ=0$ has simple exact poles when  $\omega=0$ or $q=0$:
\begin{enumerate}
\item
When $\omega=0$, 
\begin{subequations}
%\label{eq:}
\begin{align}
\calQ &= (2q^2+\eps^2)(6q^2+7\eps^2)~.\\
\to 
q^2 &=-\half \eps^2~,
 \quad(\text{universal skip})~. \\
 q^2 &=-\frac{7}{6} \eps^2~,
 \quad(\text{skip 6})~. 
%\label{eq:}
%
\end{align}
\end{subequations}

\item
When $q=0$,
%\margin{corrected 5/22}
\begin{subequations}
%\label{eq:}
\begin{align}
\calQ &= -(2\omega+i\eps^2)(60\omega+7i\eps^2)~. \\
\to
\omega &= -\frac{7i}{60}\eps^2~, 
 \quad(\text{skip 1})~,
\\
\omega &= -\half i\eps^2~,
\quad(\text{skip 7})~. 
%\label{eq:}
%
\end{align}
\end{subequations}

\end{enumerate}

%%%%%%%%%
\section{Other examples}\label{sec:further}
%%%%%%%%%

The universal pole-skipping points have the following 2 properties:
\begin{enumerate}
\item
The pole-skipping point $\nw_*=-in$ does not depend on the system one considers although $\nq_*$ may depend on the system.
\item
The pole-skipping point  $\nw_*=-in$ does not change under the higher-derivative corrections or $\alpha'$-corrections \cite{Roberts:2014isa,Grozdanov:2018kkt,Natsuume:2019vcv}.
\end{enumerate}
These properties are unlikely to hold for nonuniversal pole-skipping points. 
In order to see features of universal and nonuniversal pole-skipping points, it is useful to see some other examples.
%We briefly make remarks about some extensions  of our system.

%%------------------
\subsection{Minimal holographic \SF\ with background source}%\label{sec:}
%%------------------
We consider the solution with spontaneous condensate background $|\psi|=\eps$, but one can obtain the solution with the background source $J_0$ for $\psi$ (\appen{J}).
One can show that 
\begin{enumerate}
\item
The universal hydrodynamic pole-skipping point does not depend on $J_0$
\item
Nonuniversal hydrodynamic pole-skipping points $(\omega_*, q^2_*)$ depend on $J_0$.
\end{enumerate}

%%------------------
\subsection{Nonminimal \HSFs}%\label{sec:}
%%------------------
Second, consider a class of ``nonmimimal" \HSFHSCs:
\begin{subequations}
\label{eq:nonminimal}
\begin{align}
S_\text{m} &= -\frac{1}{g^2} \int d^5x \sqrt{-g} \biggl\{ \frac{1}{4}F_{MN}^2 + K|D_M\Psi|^2+V \biggr\}~,\\
K &= 1+A|\Psi|^2~,\quad
V= m^2 |\Psi|^2+B |\Psi|^4~.
%\label{eq:}
%
\end{align}
\end{subequations}
Here, $A$ and $B$ are bulk parameters. The original \HSFHSC\ is the $A=B=0$ case which we call the ``minimal" \HSFHSC. The arbitrary values may not be allowed for $A$ and $B$. If $A<0$, $\Psi$ may become a (unitary) ghost. If $B<0$, the potential may not be bounded below. It would be safe to set $A,B>0$.

A similar analysis can be done for nonminimal \HSFHSCs\ (\appen{nonminimal}), and the universal hydrodynamic pole-skipping point is given by
\begin{align}
(\omega_*, q^2_*)=\left(0,-\half\eps^2 \right)~.
%\label{eq:}
%
\end{align}
Namely, the pole-skipping point is the same as the minimal case, so $\omega_*=0$ does not depend on the system. (In this case, even $q_*$ does not change.)
However, one can show that most nonuniversal pole-skipping points $\omega_*$ depend on $A$ and $B$, so nonuniversal pole-skipping points are system dependent.

It would be also interesting how higher-derivative corrections change nonuniversal pole-skipping points. An analytic computation is possible even under some higher-derivative corrections \cite{2nd,2nd2}.

%%------------------
\subsection{Generic model F universality class}%\label{sec:}
%%------------------
From the dual GL theory point of view, \SFs\ are described by model F equations. One can obtain the hydrodynamic pole-skipping point for a generic model F universality class. One can rewrite the universal hydrodynamic pole-skipping point by transport coefficients as (\appen{F})
\begin{align}
(\omega_*,q_*^2) = \left(0,  -g_0^2\frac{ \tau_0+\tau_0^*}{\lambda} \eps^2 \right)~.
%\label{eq:}
%
\end{align}
Here, $\tau_0$ is the relaxation time for $\psi$ and $\lambda$ is the conductivity. Namely, the existence of the hydrodynamic pole-skipping point is generic for model F universality class. Note that $q_*$ depends on two relaxation times of the system. It would be interesting to study the other examples if this is always the case. 
%\margin{E OK}

%%%%%%%%%
\section{Discussion}%\label{sec:}
%%%%%%%%%

It would be worthwhile to study various aspects of nonuniversal pole-skipping points: 
\begin{enumerate}

\item
%However, the physical meaning of nonuniversal pole-skipping points is unclear. 
%There is a possibility that  
The origin of nonuniversal pole-skipping points is not entirely clear. 
The Green's functions with many fields are complicated in general so that they may be accidental ones. It is desirable to study  nonuniversal pole-skipping for the other systems. If they are not accidental, the ambiguity of the Green's functions should come from the ambiguity of the bulk solution.

\item
%In the SAdS$_5$ background, a field equation typically has regular singular points at $u=0,\pm 1,\infty$. 
To analyze the ambiguity of the bulk solution, one usually constructs the bulk solution as a power series from the horizon $u=1$, but this is not necessary. In principle, one can construct the full solution as%
\footnote{One should choose $u_0$ so that both $u=0$ and $u=1$ lie inside the convergence radius.}
%\footnote{The field equation in the SAdS$_5$ background usually has regular singular points at $u=0,\pm 1,\infty$. Because one needs the asymptotic form of a solution as $u\to0$, the power-series expansion is guaranteed to converge only inside a radius $\rho$ around $u=0$ in the complex $u$-plane, where $\rho$ is the distance to the nearest singular point. Because we construct the solution from $u=1$ to $u=0$, one needs $\rho \geq1$. so the power-series expansion is guaranteed to converge at $u=0$.}
\begin{align}
\phi(u) = \sum_{n=0}\phi_n(u-u_0)^{n}~.
%\label{eq:}
%
\end{align}
Because $u_0$ is an ordinary point, one can expand the solution as a Taylor series. 
One may be able to find nonuniversal pole-skipping points from such an expansion without using the full bulk solution.

\item
Also, we obtain the analytic Green's functions only in the hydrodynamic limit. It would be interesting to study pole-skipping   beyond the hydrodynamic limit to see if nonuniversal pole-skippings appear in general. 

\item
We saw that a nonuniversal pole-skipping point appears for $\nw \geq +0i$ even though our system is a Maxwell-scalar system. Universal pole-skipping points are located at $\nw \leq +i$ for bulk fields with spin-2 and lower. 
In particular, the gravitational sound mode has a universal pole-skipping point at $\nw =+i$. It would be interesting if the sound mode has a nonuniversal pole-skipping point at $\nw > +i$.
This is because this universal pole-skipping point is related to quantum many-body chaos. 

From the analysis of out-of-ordered correlators (OTOC), the quantum Lyapunov exponent has the bound \cite{Maldacena:2015waa}
\begin{align}
\frac{\lambda}{2\pi T} \leq +1~.
%\label{eq:}
%
\end{align}
The OTOC $C(t)$ then gives an exponential growth:
\begin{align}
C(t) \simeq e^{\lambda t}~.
%\label{eq:}
%
\end{align}
It is conjectured that a holographic system saturates the bound, or a black hole is maximally chaotic.
The $\nw =+i$ universal pole-skipping point in the plane-wave form coincides with this behavior $e^{-i\omega t} \sim e^{(2\pi T)t}$.

Thus, a nonuniversal pole-skipping point at $\nw > +i$ would be interesting. However, one should keep in mind that a pole-skipping point does not necessarily imply the chaos. The relation between pole-skipping phenomena and the chaos is still unclear.

\end{enumerate}

%%%%%%%%%
\section*{Acknowledgments}%\label{sec:} % for PTEP,PRD
%%%%%%%%%

%%%%%%%%%
%\acknowledgments % for JHEP
%%%%%%%%%

%\begin{acknowledgments}
I would like to thank Takashi Okamura for his continuous suggestions and interest throughout the work.
This research was supported in part by a Grant-in-Aid for Scientific Research (25K07291) from the Ministry of Education, Culture, Sports, Science and Technology, Japan. 
%\end{acknowledgments}

%%%%%%%%%

\appendix

%%%%%%%%%
\section{Supplementary information}%\label{sec:}
%%%%%%%%%

%%------------------
\subsection{The \SF\ solution by original variables}\label{sec:original}
%%------------------
For the \HSF\ in original variables, the expectation values are
%\margin{corrected 10/12}
\begin{subequations}
\label{eq:bulk_Green}
\begin{align}
\dJt &= 2q^2\frac{\calQ}{\calP} \dcalA_t + 2q\omega\frac{\calQ}{\calP} \dcalA_x+ 4\eps \frac{\calQ_{t}}{\calP} \dJ + 4\eps \frac{\calQ_{t}^*}{\calP} \dJ^*~, 
\quad O(l^2)
\\
\dJx &= 2q\omega \frac{\calQ}{\calP} \dcalA_t + 2\omega^2\frac{\calQ}{\calP} \dcalA_x+ q\eps \frac{\calQ_{x}}{\calP} \dJ - q \eps \frac{\calQ_{x}^*}{\calP} \dJ^*~,
\quad O(l^3)
\\
\vevpsi &= 4\eps \frac{\calQ_{t}}{\calP} \dcalA_t + q\eps\frac{\calQ_x}{\calP} \dcalA_x + 2\frac{\calQ_s}{\calP} \dJ
-2\eps^2\frac{Q_c}{\calP} \dJ^*~,
\quad O(l)
\\
%-\frac{2\eps^2(2q^2-16i\omega+\eps^2)}{\calP} \Jc~,\\
\vevpsic &= 4\eps \frac{\calQ_{t}^*}{\calP} \dcalA_t - q\eps\frac{\calQ_x^*}{\calP} \dcalA_x
-2\eps^2\frac{Q_c}{\calP} \dJ
%-\frac{2\eps^2(2q^2-16i\omega+\eps^2)}{\calP} J  
+2\frac{\calQ_s^*}{\calP} \dJ^*~,
\quad O(l)
%\label{eq:}
%
\end{align}
\end{subequations}
where Green's functions have a common denominator
\begin{align}
\calP &= 12q^6-48iq^4\omega-168q^2\omega^2+240i\omega^3+8q^4\epsilon^2-18iq^2\omega\epsilon^2-28\omega^2\epsilon^2+q^2\epsilon^4~,
%\label{eq:}
%
\end{align}
and there are 5 types of numerators:
\begin{subequations}
%\label{eq:}
\begin{align}
\calQ &= 12q^4 - 24iq^2\omega -120\omega^2 + 20q^2\eps^2 - 74i\omega \eps^2 + 7\eps^4~,
\\
\calQ_t &=6q^4+24(1-i)q^2\omega-60i\omega^2+3q^2\eps^2+7\omega\eps^2~,
\\
\calQ_x &=12q^4 + 24(1-2i)q^2\omega - (72+96i)\omega^2 + 8q^2\eps^2 + (12-18i)\omega\eps^2 + \eps^4~,
\\
\calQ_s &=24 q^4 + 72(1-i)q^2\omega- (48+144i)\omega^2 + 14q^2\eps^2 + (24-16i)\omega\eps^2 + \eps^4~,
\\
\calQ_c &=2q^2-16i\omega+\eps^2~.
%\label{eq:}
%
\end{align}
\end{subequations}
We work in the momentum space (Fourier space). So, a complex conjugate solution is obtained by $\omega\to-\omega, q\to-q$ and taking the complex conjugate. $\omega$ appear in the form of $w := i\omega$, so take the complex conjugate as if $w$ is real.%
\footnote{We work in the momentum space, so $\psi$ here is $\psi_k$, where $\psi(t,\bmx,u)=\int (dk) e^{-i\omega t+i \bmq\cdot\bmx}\psi_k(u)$ with $(dk)=d^4k/(2\pi)^4$. Then, the complex conjugate field is given by $\psi_{-k}^*$.
}

There are 16 Green's functions, but
\begin{enumerate}
\item
They are symmetric $G^{ij}=G^{ji}$. This leaves 10 Green's functions.
 %satisfy Onsager's reciprocal relations
\item
In addition, the Ward identity hold in the Maxwell sector. This leaves 8 Green's functions.

\item
Also, the conservation law $\del_\mu\bra \delta\calJ^\mu \ket=i\epsilon (\dJ-\dJ^*)$ leaves 6 independent Green's functions.
%\begin{align}
%
%(-i\omega G^{ts}+iq G^{xs})J = -i\eps J~,
%(-i\omega G^{tc}+iq G^{xc})\Jc = i\eps \Jc~.
%\label{eq:}
%
%\end{align}
%As a result,
%\begin{align}
%
%q\eps\frac{\calQ_x}{\calP} = \frac{\eps}{q} \left(1+4\omega \frac{\calQ_t}{\calP} \right)
%\label{eq:}
%
%\end{align}
%\item
The independent numerators are $Q, Q_t,Q_t^*,Q_s, Q_s^*, Q_c$.
\end{enumerate}

%%------------------
\subsection{Green's functions near nonuniversal points}\label{sec:slope}
%%------------------
In this appendix, we list Green's functions near nonuniversal pole-skipping points.

%%------------------
\subsubsection{Holographic \SF}%\label{sec:}
%%------------------
%\paragraph{Holographic \SF:} 
A nonuniversal pole-skipping point appears only for some Green's functions. So, we do not have to fix how one approaches the pole-skipping point. One could choose, \eg, $\delta q$ or $\dqs$. But we keep using $\dw,\dqs$.
Also, we omit $G^{\rho t},G^{\theta t}, G^{\theta\rho}$ because they are not independent. For Case 2 and 3 (\sect{def}), we suppress all numerical coefficients.
%\begin{subequations}
%\label{eq:}
%\begin{align}
%
%G^{tt} &\sim O \left(\frac{\dqs(\dqs-\dw)}{\dqs-\dw} \right) \to 0~,1%0714
%\\
%G^{xt}&\sim O(\dq)?~,1
%\label{eq:}
%
%\end{align}
%\end{subequations}

%\margin{checked}
%\begin{enumerate}[(a)]
\begin{itemize}

\item
skip 0: $(\omega,q^2)=(0,0)$:
\begin{subequations}
%\label{eq:}
\begin{align}
\dJt &\sim
14 \dmfA_t
+\frac{12}{\eps}\dJr
+\frac{28}{\eps}\frac{ i \dw}{\dqs}\dJi %0714
~,
%0
\\
\dJx  &\sim
O\left(\frac{\dw}{\sqrt{\dqs}} \right)\times \dmfA_t
+O\left(\frac{\dw}{\sqrt{\dqs}} \right) \times\dJr
+O\left(\frac{1}{\sqrt{\dqs}} \right) \times\dJi
% +\frac{i  \eps}{ \sqrt{\dqs}}\dJi%0714
 ~,%0
\\
\drho &\sim 
\frac{12}{\eps} \dmfA_t
+\frac{12}{\eps^2 }\dJr
+\frac{24}{\eps^2}\frac{ i \dw }{\dqs}\dJi %0714
~.%0
%\braket{\delta\itheta} & \sim
% -\frac{28}{\eps}\frac{ i \dw}{\dqs} \dmfA_t
% -\frac{24}{\eps^2}\frac{i \dw }{\dqs}\dJ_\rho
%  +\frac{2}{\dqs}\dJ_\theta (\infty) %0714
%  ~,0 
%\\
%\label{eq:}
%
\end{align}
\end{subequations}
This is not a pole-skipping point for $G^{\theta\theta}$.
$G^{xt}, G^{x\rho}$ vanishes because we take $\dw \sim \dqs$ (Case 2).
$G^{x\theta}$ diverges but corresponds to Case 3. 
%We omit all numerical coefficients for those Green's functions.

\item
skip 1: $\left(-\dfrac{7i}{60}\eps^2,0 \right)$
\begin{subequations}
%\label{eq:}
\begin{align}
\dJt &\sim
O(\dqs)\times \dmfA_t
 + \frac{60}{\eps}\frac{\dqs }{89\dqs-245 i\dw}\dJr
+ \frac{420}{\eps}\frac{ 2\dqs-5 i\dw}{89\dqs-245 i\dw}\dJi~, %1
   \\
\dJx &\sim 
O(\sqrt{\dqs})\times \dmfA_t
+O \left(\frac{\sqrt{\dqs}}{\dqs-i\dw} \right) \times\dJr
%-7i \eps\frac{\sqrt{\dqs}}{89\dqs-245i\dw}\dJ_\rho
+O \left(\frac{\sqrt{\dqs}}{\dqs-i\dw} \right) \times\dJi
%-9i \eps\frac{\sqrt{\dqs}}{89\dqs-245i\dw}\dJ_\theta
 ~. %1
%G^{t\rho} &\sim \frac{60}{\eps} \frac{\dqs}{245i\dw-89\dqs}~, 1\\
%G^{t\theta} &\sim -\frac{420}{\eps} \frac{5i\dw-2\dqs}{245i\dw-89\dqs}~, 1\\
%G^{x\rho} &\sim -\frac{\eps}{35}\frac{\dq}{\dw}~, 1\\
%G^{x\theta} &\sim -\frac{9\eps}{245}\frac{\dq}{\dw}~,1 
%\label{eq:}
%
\end{align}
\end{subequations}
$G^{tt},G^{xt}$ vanish (Case 2). $G^{x\rho},G^{x\theta}$ diverge but correspond to Case 3. 
%In addition, $G^{tt}$ does not have the slope dependence and is constant (Case 1).
%$G^{xt} \to 0$ (Case 2).

\item
skip 2: $\left(-\dfrac{5i}{36}\eps^2,\dfrac{1}{18}\eps^2 \right) $
\begin{subequations}
%\label{eq:}
\begin{align}
\dJt &\sim 
  \dmfA_t
 +\frac{2}{\eps}\frac{\dqs-4 i\dw}{3\dqs-7 i\dw}\dJr 
+\frac{10}{\eps}\frac{2\dqs-5 i\dw}{3\dqs-7 i\dw}\dJi~,%2
   \\
%\dJx & \sim 
%5 \eps \frac{2\eps\dq -7 i \sqrt{2}\dw}{84\dw+12 i \sqrt{2}\eps\dq } \dmfA_t 
%  +\frac{ 30 \sqrt{2} \dw+5 i\eps\dq }{-9 \sqrt{2}\eps\dq +63 i\dw}\dJ_\rho
%+\frac{3 \sqrt{2}\dw+8 i\eps\dq  }{18 \left(\sqrt{2}\eps\dq -7 i\dw\right)}\dJ_\theta~,2 % dq
%\\
\dJx &\sim 
-\frac{5 i\eps}{6 \sqrt{2}} \dmfA_t
-\frac{5i}{3\sq}\frac{\dqs-4 i \dw}{3\dqs-7 i \dw}\dJr
+\frac{1}{3\sq}\frac{\dw+4 i \dqs}{3 \dqs-7 i\dw}\dJi %0714
~. %2
%G^{t\rho} &\sim -\frac{2}{\eps} \frac{4i\dw-\dqs}{7i\dw-3\dqs}~, 2\\
%G^{t\theta} &\sim -\frac{10}{\eps} \frac{5i\dw-2\dqs}{7i\dw-3\dqs}~, 2\\
%G^{x\rho} &\sim \frac{60i\dw-5\sq\eps\dq}{63\sq \dw+18i\eps\dq}~, 2\\
%G^{x\theta} &\sim- \frac{i}{9\sq}\frac{3i\dw-4\sq\eps\dq}{7i\sq \dw-\sq\eps\dq}~, 2
%\label{eq:}
%
\end{align}
\end{subequations}
%In addition, $G^{tt},G^{xt}$ do not have the slope dependence and are constant (Case 1).

\item
skip 3: $\left(-\dfrac{i}{12}\eps^2,-\dfrac{1}{6}\eps^2 \right) $
\begin{subequations}
%\label{eq:}
\begin{align}
G^{t\rho} &\sim \frac{12}{\eps} \frac{4i\dw-\dqs}{10i\dw-\dqs}~, %3
\\
G^{x\rho}& \sim
 -\sqrt{6}\frac{\dqs-4 i \dw}{\dqs-10 i \dw} %0714
 ~, %3
\\
%G^{x\rho} &\sim -6\frac{2\sqrt{6}\dw-\eps\dq}{30 \dw-\sqrt{6}\eps\dq}~, 3\\ % dq
%G^{\rho\rho} &\sim (\text{constant})~, \\
G^{\rho\theta} &\sim -\frac{18}{\eps^2}\frac{2i\dw-\dqs}{10i\dw-\dqs}~. %3
%\label{eq:}
%
\end{align}
\end{subequations}
In addition, $G^{\rho\rho}\sim 6/\eps^2$ (Case 1).

\item
skip 4: $\left(0,-\dfrac{1}{6}\eps^2 \right)$
\begin{subequations}
%\label{eq:}
\begin{align}
G^{t\theta} &\sim -\frac{36}{\eps} \frac{i\dw}{5i\dw-2\dqs}~, %4
\\
\dJx &\sim 
 -\frac{4 \sqrt{6} \eps i\dw }{2 \dqs-5 i\dw} \dmfA_t
 -\frac{4 \sqrt{6} i\dw}{2 \dqs-5 i\dw}\dJr
 +\sqrt{6}\dJi %0714
 ~,%4
 \\
%G^{xt} &\sim -\frac{12\sqrt{6}\eps\dw}{15 \dw-2\sqrt{6}\eps\dq}~, 4\\
%G^{x\rho} &\sim 4\sqrt{6}\frac{i\dw}{2\dqs-5i\dw} %0714
%~,4
%\\
%G^{x\rho} &\sim \frac{12\sqrt{6}\dw}{-15 \dw+2\sqrt{6}\eps\dq}~, 4\\ %dq
G^{\rho\theta} &\sim -\frac{36}{\eps^2}\frac{i\dw}{5i\dw-2\dqs}~. %4
%\label{eq:}
%
\end{align}
\end{subequations}
In addition, $G^{\theta\theta}\sim 12/\eps^2$ (Case 1).

\item
skip 5: $\left(+\dfrac{i}{12}\eps^2,-\dfrac{1}{2}\eps^2 \right) $ 
\begin{subequations}
%\label{eq:}
\begin{align}
G^{t\theta} &\sim -\frac{6}{\eps} \frac{5i\dw-2\dqs}{2i\dw+\dqs}~, %5
\\
G^{x\theta} &\sim -\frac{9}{\sq}\frac{i\dw}{\dqs+2 i \dw} %0714
~,%5
\\
%G^{x\theta} &\sim -\frac{9\dw}{2\sq\dw+2\eps\dq}~, 5 \\ %dq
G^{\rho\theta} &\sim -\frac{9}{\eps^2}\frac{2i\dw-\dqs}{2i\dw+\dqs}~. %5
%\label{eq:}
%
\end{align}
\end{subequations}
In addition, $G^{\theta\theta}\sim 3/\eps^2$ (Case 1).

\end{itemize}
%\end{enumerate}

%%------------------
\subsubsection{Holographic \SC}%\label{sec:}
%%------------------
%\paragraph{Holographic \SC:}
%\margin{checked}
%$A^{tx} \sim -\frac{2\dw}{\eps^2 \dq}~, 0?$

%\begin{enumerate}[(a)]
\begin{itemize}

\item
skip 1:
$\left(-\dfrac{7i}{60}\eps^2,0 \right)$
%\begin{align}
%
%G^{\rho x} &\sim \frac{21\dq}{115\eps \dw}~, 1
%\label{eq:}
%
%\end{align}
\begin{subequations}
%\label{eq:}
\begin{align}
A^{tx} &\sim O\left( \frac{1}{\sqrt{\dqs}} \right)~, \\
\tilG^{\rho x} &\sim O \left(\frac{\sqrt{\dqs}}{\dqs-i\dw} \right)~, \\
\tilG^{\theta x} &\sim O \left(\frac{1}{\sqrt{\dqs}} \right)~.
%\label{eq:}
%A^{tx} &\sim +O \left(\frac{\sqrt{\dqs}}{\dqs-\dw} \right) \times\dJr
\end{align}
\end{subequations}
They correspond to Case 3.
%$A^{tx}, \tilG^{\rho x},\tilG^{\theta x}\to\infty$ (Case 3).

\item
skip 2:
$\left(-\dfrac{5i}{36}\eps^2,\dfrac{1}{18}\eps^2 \right) $ 
\begin{align}
%
% A^{tx} &\sim \frac{5\sq}{3\eps}\frac{5 \dw+2 i \dqs}{3 \dqs-7 i \dw}~,2 %0714
%\\
\dmfA_t &\sim 
- \frac{2}{\eps}\frac{\dqs-4 i\dw}{3\dqs-7 i\dw}\dJr
  +\frac{ -4\dqs+i\dw}{18 \{3\dqs-7 i\dw\}} \Jte
  +\frac{5\sq}{3\eps}\frac{-5\dw-2 i\dqs}{3\dqs-7 i\dw} \Jxe~. %2
%A^{t\rho} &\sim \frac{2}{\eps} \frac{4i\dw-\dqs}{7i\dw-3\dqs}~, 2 \\
%A^{tt} &\sim \frac{i\dw-4\dqs}{126i\dw-54\dqs}~, 2\\
%A^{tx} &\sim \frac{75\sq \dw +20i\eps \dq}{9(-7i\eps\dw+\sq\eps^2\dq) }~, 2
%\label{eq:}
%
\end{align}

\item
skip 6:
$\left( 0,-\dfrac{7}{6}\eps^2 \right) $
%\margin{corrected}
\begin{subequations}
%\label{eq:}
\begin{align}
A^{tx} &\sim \frac{3\sqrt{42}}{\eps}\frac{i\dw}{ 23i\dw+4\dqs }~, %6
\\
%A^{tx} &\sim \frac{9\sqrt{42}\dw}{ 69\eps \dw+4\sqrt{42}\eps^2\dq }~, 6 %dq
%\\
\tilG^{\rho x} &\sim
-\frac{6\sqrt{42}}{\eps^2}\frac{i\dw}{23i\dw+4\dqs}~, %6
\\
%G^{\rho x} &\sim
%-\frac{1}{\eps^2}\frac{18\sqrt{42}\dw}{69\dw+4\sqrt{42}\eps\dq}~,6 %dq
%\\
%G^{\theta\rho} &\sim -\frac{36}{\eps^2}\frac{i\dw}{23i\dw+4\dqs}~, 6
%\\
%G^{\theta t} &\sim
%\frac{18}{\eps} \frac{\dw}{23\dw-4i\dqs}~, 6\\
\braket{\delta\itheta} &\sim 
\frac{36}{\eps^2}\frac{i\dw}{4\dqs +23 i\dw }\dJr
   -\frac{18}{\eps}\frac{\dw}{23\dw-4 i\dqs}\Jte
   +2\sqrt{\frac{6}{7}} \frac{1}{\eps^2 } \Jxe~. %6
%\label{eq:}
%
\end{align}
\end{subequations}
%In addition, $G^{\theta x}$ does not have the slope dependence and is finite (Case 1).

\item
skip 7:
$\left( -\dfrac{1}{2}\eps^2,0 \right) $
\begin{subequations}
%\label{eq:}
\begin{align}
\drho& \sim 
 \frac{1}{\eps^2}\frac{ 12\dqs-12 i\dw}{4\dqs +23 i\dw  }\dJr
   -\frac{6}{\eps}\frac{\dqs+2 i\dw}{ 4\dqs+23 i\dw } \Jte
  +O \left( \frac{\sqrt{\dqs}}{\dqs-i\dw} \right) \times \Jxe~,\\
%+\frac{18 i \sqrt{\dqs} }{\eps (4\dqs+23 i\dw)} \Jxe~, %7%\\
%G^{\rho\rho} &\sim \frac{12{{\eps^2}\frac{i\dw-\dqs}{23i\dw+4\dqs}~, 7\\
%G^{\rho t} &\sim \frac{6}{\eps} \frac{2i\dw+\dqs}{23i\dw+4\dqs}~, 7 \\
\tilG^{\rho x} &\sim O \left( \frac{\sqrt{\dqs}}{\dqs-i\dw} \right)~.
%G^{\rho x} &-\frac{18\dq}{23\eps\dw}~,7
%\label{eq:}
%
\end{align}
\end{subequations}
$\tilG^{\rho x}$ diverges but corresponds to Case 3. 

\item
skip 8:
$\left( \dfrac{-3i-\sqrt{11}}{12}\eps^2, \dfrac{7-\sqrt{11}i}{12}\eps^2 \right) $
\begin{subequations}
%\label{eq:}
\begin{align}
A^{tt} &\sim \frac{ (-5i+4\sqrt{11})\dw+(5+i\sqrt{11})\dqs }{ 2(-14i+9\sqrt{11})\dw+28\dqs }~, %8
\\
\tilG^{\rho t} &\sim
\frac{6}{\eps} \frac{(\sqrt{11}-i)\dw+\dqs}{(9\sqrt{11}-14i)\dw+14\dqs}~, %8
   \\
\tilG^{\theta t} &\sim
-\frac{3-i\sqrt{11}}{\eps} \frac{i\dw-\dqs}{ (9\sqrt{11}-14i)\dw+14\dqs}~. %8
%\label{eq:}
%
\end{align}
\end{subequations}

\item
skip 9:
$\left( \dfrac{-3i+\sqrt{11}}{12}\eps^2, \dfrac{7+\sqrt{11}i}{12}\eps^2 \right) $ 
\begin{subequations}
%\label{eq:}
\begin{align}
A^{tt} &\sim \frac{ (5i+4\sqrt{11})\dw-(5-i\sqrt{11})\dqs }{ 2(14i+9\sqrt{11})\dw-28\dqs }~, %9 
\\
\tilG^{\rho t} & \sim
\frac{6}{\eps} \frac{(\sqrt{11}+i)\dw-\dqs}{(9\sqrt{11}+14i)\dw-14\dqs}~, %9
\\
\tilG^{\theta t} &\sim
\frac{3+i\sqrt{11}}{\eps} \frac{i\dw-\dqs}{ (9\sqrt{11}+14i)\dw-14\dqs}~. %9
%\label{eq:}
%
\end{align}
\end{subequations}

\end{itemize}
%\end{enumerate}

%%%%%%%%%
\section{Other examples: some details}%\label{sec:}
%%%%%%%%%

%%------------------
\subsection{Minimal \HSF\ with background source $J_0$}\label{sec:J}
%%------------------

In the text, we consider the solution with spontaneous condensate background $|\psi|=\eps$, but one can obtain the solution with the background source $J$ for $\psi$.

For the dual GL theory, decompose $\psi$ and $J$ as the amplitudes and the phases:
\begin{align}
\psi =\rho e^{i\theta}~, \quad J=J_\rho e^{i\varphi}~,
%\label{eq:}
%
\end{align}
The gauge transformation acts as
\begin{align}
\theta \to \theta+\Lambda~, \varphi \to \varphi+\Lambda~,
%\label{eq:}
%
\end{align}
so that the source term in the free energy is gauge invariant:
\begin{align}
\psi J^*+\psi^* J=2\rho J_\rho \cos(\theta-\varphi)~.
%\label{eq:}
%
\end{align}

Write the background as
\begin{align}
\psi = \eps~, J=J_0~.
%\label{eq:}
%
\end{align}
One can set $\psi$ and $J$ to be real without loss of generality.
The perturbations are given by
\begin{align}
\delta\psi =\delta\rho+i\eps\delta\theta~,
\delta J= \delta J_\rho+iJ_0 \delta \varphi~.
%\label{eq:}
%
\end{align}
$\delta\rho,\delta J_\rho$ are gauge invariant variables, and the following variables are gauge-invariant as well:
\begin{align}
\delta\itheta =\eps \left(\delta\theta-\frac{1}{iq} \delta\calA_x \right)~,
%\delta\varphi_J
\dJ_\theta =J_0 \left(\delta\varphi-\frac{1}{iq} \delta\calA_x \right)~.
%\label{eq:}
%
\end{align}
Here, $\delta \calA_x \to \delta \calA_x +iq \Lambda$. $\dJ_\theta$ can be written as
\begin{align}
%
%\delta\varphi_J
\dJ_\theta=\frac{\delta J-\delta \Jc}{2i}-\frac{J_0}{iq} \delta\calA_x~.
%\label{eq:}
%
\end{align}
$\dJ_+=2\dJ_\rho, \dJ_-=2\dJ_\theta$.
%\delta\varphi_J$.

It is straightforward to construct the bulk solution, and the solution is written by gauge-invariant variables, but the solution is rather cumbersome, so we omit its explicit form. The $J_0=0$ results in the text can be regarded as the $J_0=0$ limits.  One can discuss pole-skipping points in this case as well. One can show that the universal hydrodynamic pole-skipping point does not depend on $J_0$ whereas nonuniversal hydrodynamic pole-skipping points typically depend on $J_0$. 

One can show that the bulk solution agrees with model F equations with $J_0$ (\appen{F}). In this case, the background \eqref{eq:F_back} is replaced by
\begin{subequations}
%\label{eq:}
\begin{align}
\rho_0 &= \half \eps^2~,\\
a \eps&=\frac{1}{48} \eps^3-J_0~.
%\label{eq:}
%
\end{align}
\end{subequations}
It is straightforward to solve $\dpsi,\dpsi^*,\delta \rho$ equations in the background.

%%------------------
\subsection{Nonminimal \HSFs}\label{sec:nonminimal}
%%------------------

The solution is given by
\begin{subequations}
\label{eq:nonminimal_sol}
\begin{align}
\dJt &= 
2 q^2\frac{\calQ}{\calP} \dmfA_t 
+12 \eps \frac{\Qt}{\calP}\dJr
+ 4\eps \frac{\Qtp}{\calP}\dJi~, 
\quad O(l^2)
\\
\dJx &= 
\frac{\omega}{q} \left(2q^2\frac{\calQ}{\calP} \dmfA_t
+ 12\eps \frac{\Qt}{\calP}\dJr
+4\eps\frac{\Qtp}{\calP} \dJi \right) 
- \frac{\eps}{iq}\dJi
\quad O(l^3)
\\
\drho &= 
12\eps \frac{\Qt}{\calP} \dmfA_t 
+ 12 \frac{\Qs}{\calP}\dJr
+24\frac{\Qsp}{\calP}\dJi~, 
\quad O(l)
\\
\braket{\delta\itheta} &= 
-4\eps \frac{\Qtp}{\calP} \dmfA_t 
-24 \frac{\Qsp}{\calP}\dJr
+2\frac{\Qi}{\calP} \dJi~, 
\quad O(l)
%\label{eq:}
%
\end{align}
\end{subequations}
The solution has a common denominator
\begin{align}
\calP &= 12q^6 %
-48iq^4\omega %
-168q^2\omega^2 %
+240i\omega^3 %
+8(1+A+B)q^4\epsilon^2 %
-6(3+4A+4B) q^2i\omega\epsilon^2 %
\nonumber \\
&-4(7+4A+4B) \omega^2\epsilon^2 %
+(1+4A+4B) q^2\epsilon^4~,
%\label{eq:}
%
\end{align}
and there are 6 types of $O(l^4)$ numerators:
%\margin{checked}
%\margin{corrected 9/10}
\begin{subequations}
%\label{eq:}
\begin{align}
\calQ &= 12q^4 %
 - 24iq^2\omega %
 -120\omega^2 %
 + 4(5+2A+2B) q^2\eps^2 %
 -2(37+4A+4B)i\omega \eps^2 %
 \nonumber \\
& + (7+4A+4B)\eps^4~, %
\\
\Qt &= q^2(2q^2-8i\omega+\eps^2)~, %
\\
\Qtp &= i\omega\{24q^2-60i\omega+(7+4A+4B)\eps^2\}~,%
\\
%\calQ_x &=12q^4 -48i\omega q^2-72\omega^2+8q^2\eps^2-18i\omega\eps^2+\eps^4~,
%\\
\Qs &=2q^4 -6iq^2\omega -4\omega^2 +q^2\eps^2~, %
\\
\Qsp &=i\omega(3q^2-6i\omega+\eps^2)~,%
\\
\Qi &=12q^4%
-36iq^2\omega %
-24\omega^2 %
+8(1+4A+4B)q^2\eps^2
-16(1+4A+4B) i\omega\eps^2
\nonumber \\
&+ (1+4A+4B)\eps^4~.
%\label{eq:}
%
\end{align}
\end{subequations}
The current is conserved when there are no scalar source $\dJi$:
\begin{align}
\del_\mu\braket{\delta\calJ^\mu}= -\eps\dJi~,
\quad O(l^4)~.
%\label{eq:}
%
\end{align}

It is straightforward to obtain approximate dispersion relations and pole-skipping points. By imposing the Neumann boundary condition, one can obtain the results for ``nonminimal" \HSCs.

%\newpage
%%------------------
\subsection{The hydrodynamic pole-skipping for model F universality class}\label{sec:F}
%%------------------

We discuss hydrodynamic pole-skipping points for a particular \HSFHSC\ in the text, but the existence of a universal hydrodynamic pole-skipping point itself is generic for an arbitrary bulk scalar mass $m^2$ \cite{Natsuume:2025nkx}. However, one cannot write the results by boundary quantities such as $\eps$ because the analytic solution is unavailable. 

Meanwhile, we have shown that our solution is equivalent to model F equations \cite{4th}. One can obtain the hydrodynamic pole-skipping point for a generic model F universality class. In this case, one can rewrite the pole-skipping point by transport coefficients as we see below. 

%The dynamic universality classes were classified by Hohenberg and Halperin \cite{hohenberg_halperin}. Among them, a \SF/Type II \SC\ belong to model F universality class. 
In model F, a non-conserved order parameter is coupled to the other conserved modes. For our \HSF, they correspond to the  ``macroscopic wave function"  $\psi$ and the charge density $\rho$, respectively.%
\footnote{$\rho$ in this section represents the charge density, not the amplitude of $\psi$.}
%The order parameter is the ``macroscopic wave function" $\psi$ that is not conserved. $\psi$ couples to a conserved charge density $\rho$ in the low-temperature phase (ordered phase).

%We would like to identify the dual GL theory from bulk results in \sect{bulk}.
%The scalar mode can be explained by model F equations.
Our model F equations  \cite{4th} are given by
\begin{subequations}
\label{eq:F}
\begin{align}
D_t\psi &= - \tau_0^{-1}\frac{\delta F}{\delta \psi^*}  -ig_0 \psi\frac{\delta F}{\delta \rho}~, \\
\del_t \rho &= \lambda \del_i^2 \left( \frac{\delta F}{\delta \rho} \right) - \lambda \del_iE^i  +2g_0 \Im \left( \psi^*\frac{\delta F}{\delta \psi^*} \right)~,
%\label{eq:}
%
\end{align}
\end{subequations}
as well as the complex conjugate equation for $\psi^*$. 
Here, $D_\mu=\del_\mu-ig_0 \calA_\mu$ and $E^i$ is the electric field $E^i=\calF_{it} =\del_i \calA_t- \del_t \calA_i$. The GL free energy $F=\int d^3x f$ is given by
\begin{align}
f &= c_K|D_i\psi|^2 - a |\psi|^2 + \frac{b'}{2}|\psi|^4+
\frac{1}{2C}\rho^2 +\gamma \rho |\psi|^2
%-\rho \At 
-(\psi J^*+\psi^* J)~. 
\end{align}
We identify $\rho$ as the charge density and $\calA_t$ as the gauge potential. The $\rho$-equation represents the current conservation law.
$J,J^*$ represent the source terms for $\psi, \psi^*$. 
%In real experimental situations, one is not able to have the control parameters for macroscopic wave functions, but in principle, one is able to consider these source terms according to the linear response theory.
% ($\rho$ in this section is not gauge-invariant variable).

\begin{enumerate}
\item
Because we identify model F as the one with Maxwell field, we slightly modify model F equations so that the system includes both $\calA_t$ and $\calA_i$. We also write down the equations so that the $U(1)$ gauge symmetry is manifest.
%\begin{enumerate}
%\item
%For the standard model F, the free energy has the $\rho \calA_t$ coupling and becomes the covariant derivative $D_t$. We omit this term from $f$ and include it in the equation of motion. 
%\item
%More importantly, the standard $\rho$-equation has only $\calA_t$ and does not couple to $\calA_i$. (For a \HSC, the vector potential should exist. For a holographic superfluid, it corresponds to a superfluid flow.) As a result, the conventional model F equation is not gauge-invariant and is valid only when $\calA_i=0$.
%\end{enumerate}
%\item
%In the stationary case without the gauge potential $\calA_t$, one obtains the standard GL theory:
%\margin{TBC}
%\begin{align}
%
%0= \frac{\delta F}{\delta \psi^*} = c_K D_i^2 \psi-a\psi-b'\psi|\psi|^2 +J~.
%\label{eq:}
%
%\end{align}

\item
The model F has 5 additional parameters $\tau_0, D,C,g_0,\gamma$ in addition to the stationary case parameters $c_K,a,b'$.
$\tau_0$ represents the relaxation time for $\psi$ although $\tau_0 \in \mathbb{C}$, $C$ is the thermodynamic susceptibility, $\lambda$ is the conductivity, and $D:=\lambda/C$ is the diffusion constant for $\rho$. 
\item
The coefficient $c_K$ is actually redundant because one can always absorb it by a $\psi$ scaling. Thus, there are 2 independent parameters $a,b'$. But it is useful to keep it to compare with holographic results. This is because the dual GL theory typically does not have the canonical normalization if one uses the standard AdS/CFT dictionary. 

\end{enumerate}
%Namely, we prefer that the dissipative term, which breaks the time reversal symmetry, are added on the left-hand side of equations. 
%\margin{E}

The model F equations are equivalent to our minimal \HSF\ in the hydrodynamic limit with the following set of coefficients \cite{4th}: 
\begin{align}
\tau_0=\frac{1-3i}{4},
\quad
D=\half, 
\quad
C=2, 
\quad
g_0=1,
\quad
\gamma=-\frac{1}{4},
\quad
c_K=\frac{1}{4}~,
\quad
a= \half \epsmu,
\quad
b'=\frac{7}{48}~.
%\label{eq:}
%
\end{align}
This was shown when the scalar source background $J_0=0$, but this can be shown when $J_0\neq0$ as well. For nonminimal \HSFs\ (\appen{nonminimal}), the only difference is $b'=(7+4A+4B)/48$.

\eq{F} become
\begin{subequations}
%\label{eq:}
\begin{align}
D_t\psi &= \tau_0^{-1}[ c_K D_i^2\psi + a\psi - b'\psi|\psi|^2 - \gamma \rho\psi +J] -ig_0 \psi\left( \frac{1}{C} \rho + \gamma|\psi|^2\right)~, \\
%\tau_0 \del_t\psi &= c_K \del_i^2\psi+ \left(a-\gamma m \right)\psi-\frac{ig}{C}m\psi-(b+ig\gamma)\psi|\psi|^2 +J + ig\psi h \\
\del_t \rho &= \lambda \del_i^2 \left( \frac{1}{C} \rho + \gamma|\psi|^2 - \dmfA_t \right) +2g_0 \Im \left( \psi^*\frac{\delta F}{\delta \psi^*} \right)~.
%\label{eq:}
%
\end{align}
\end{subequations}

%\paragraph{Low-temperature phase:}
From the $\psi$-equation, the spontaneous condensate background $\eps$ ($J=0$) and the background charge density $\rho_0$ is given by
\begin{align}
0= -ig_0\eps \left(\frac{\rho_0}{C}+\gamma\eps^2 \right)+\eps\frac{a-\rho_0\gamma-b'\eps^2}{\tau_0}~.
%\label{eq:}
%
\end{align}
The $\psi^*$ equation gives the complex conjugate equation, and $\rho_0$ cannot vanish. The real part and the imaginary part must vanish separately:
%\margin{TBC}
\begin{subequations}
\label{eq:F_back}
\begin{align}
0&=\frac{\rho_0}{C}+\gamma\eps^2 \to \rho_0 = - C\gamma\eps^2~,\\
0&=a - \rho_0\gamma - b'\eps^2 \to b'=\frac{a}{\eps^2}+C \gamma^2~.
%\label{eq:}
%
\end{align}
\end{subequations}

Consider the linear perturbations around the background:
\begin{align}
\psi =\eps+\dpsi~,
\psi^* =\eps+\dpsi^*~,
\rho =\rho_0+ \delta\rho~.
%\label{eq:}
%
\end{align}
It is straightforward to solve $\dpsi,\dpsi^*,\delta \rho$ equations in terms of sources $\dmfA_t, \dJ, \dJ^*$. 
One can then find the universal hydrodynamic pole-skipping point:
\begin{align}
(\omega,q^2) = \left(0,  -g_0^2\frac{ \tau_0+\tau_0^*}{\lambda} \eps^2 \right)~.
%\label{eq:}
%
\end{align}
For our \HSF, $\tau_0=(1-3i)/4, \lambda=g_0=1$, so the result reduces to
\begin{align}
(\omega,q^2)= \left(0, -\half\eps^2 \right)~,
%\label{eq:}
%
\end{align}
that agrees with the holographic result. 

How about the hydrodynamic pole-skipping point for the other dynamic universality classes? 
\begin{enumerate}
\item
For model E universality class, $\gamma=0$, and equations are invariant under $\rho\to-\rho,\psi\to\psi^*$. In this case, there is no hydrodynamic pole-skipping point.
\item
For model A universality class, a non-conserved order parameter decouples from the conserved mode. One example is the \HSF\ in the high-temperature phase, where the Maxwell field decouples. In this case, there is no hydrodynamic pole-skipping point as in \eq{high}.
\end{enumerate}
It would be interesting to study the other dynamic universality classes.

\footnotesize

%\newpage

\end{document}